\pdfoutput=1
\documentclass{JINST}

\usepackage{graphicx}
\usepackage{amsmath}

\title{Hadronic energy resolution of a highly granular scintillator-steel hadron calorimeter using software compensation techniques}

\author{\centering 
\LARGE\bf The CALICE Collaboration
}

\author{\centering
C.\,Adloff, 
J.\,Blaha, 
J.-J.\,Blaising, 
C.\,Drancourt,
A.\,Espargili\`{e}re, 
R.\,Gaglione, 
N.\,Geffroy, 
Y.\,Karyotakis, 
J.\,Prast,
G.\,Vouters
\\ \it
Laboratoire d'Annecy-le-Vieux de Physique des Particules, Universit\'{e} de Savoie,
CNRS/IN2P3,
9 Chemin de Bellevue BP110, F-74941 Annecy-le-Vieux CEDEX, France
}

\author{\centering
K.\,Francis,
J.\,Repond, 
J.\,Smith\footnote{Also at University of Texas, Arlington},
L.\,Xia 
\\ \it
Argonne National Laboratory,
9700 S.\ Cass Avenue,
Argonne, IL 60439-4815,
USA}

\author{\centering
E.\,Baldolemar, 
J.\,Li\footnote{Deceased}, 
S.\,T.\,Park, 
M.\,Sosebee, 
A.\,P.\,White, 
J.\,Yu
\\ \it
Department of Physics, SH108, University of Texas, Arlington, TX 76019, USA
}

\author{\centering
T.\,Buanes, G.\,Eigen
\\ \it
University of Bergen, Inst.\, of Physics, Allegaten 55, N-5007 Bergen, Norway
}

\author{\centering
Y.\,Mikami, 
N.\,K.\,Watson 
\\ \it
University of Birmingham,
School of Physics and Astronomy,
Edgbaston, Birmingham B15 2TT, UK
}
\author{\centering 
T.\,Goto, 
G.\,Mavromanolakis\footnote{Now at CERN}, 
M.\,A.\,Thomson, 
D.\,R.\,Ward, 
W.\,Yan\footnote{Now at Dept.\ of Modern Physics, Univ.\ of Science and Technology of China, 96 Jinzhai Road, Hefei, Anhui, 230026, P.\, R.\, China}
\\ \it
University of Cambridge, Cavendish Laboratory, J J Thomson Avenue, CB3 0HE, UK
}

\author{\centering 
D.\,Benchekroun, 
A.\,Hoummada, 
Y.\,Khoulaki
\\ \it
Universit\'{e} Hassan II A\"{\i}n Chock, Facult\'{e} des sciences.\, B.P. 5366 Maarif, Casablanca, Morocco
}

\author{\centering
M.\,Benyamna, 
C.\,C\^{a}rloganu, 
F.\,Fehr, 
P.\,Gay, 
S.\,Manen, 
L.\,Royer
\\ \it
Clermont Univertsit\'e, Universit\'e Blaise Pascal, CNRS/IN2P3, LPC, BP
10448, F-63000 Clermont-Ferrand, France
}

\author{\centering
G.\,C.\,Blazey,
A.\,Dyshkant, 
J.\,G.\,R.\,Lima, 
V.\,Zutshi
\\ \it
NICADD, Northern  Illinois University,
Department of Physics,
DeKalb, IL 60115,
USA
}

\author{\centering 
J.\,-Y.\,Hostachy, 
L.\,Morin
\\ \it
Laboratoire de Physique Subatomique et de Cosmologie - Universit\'{e} Joseph Fourier Grenoble 1 -
CNRS/IN2P3 - Institut Polytechnique de Grenoble,
53, rue des Martyrs,
38026 Grenoble CEDEX, France
}

\author{\centering 
U.\,Cornett, 
D.\,David, 
G.\,Falley, 
K.\,Gadow, 
P.\,G\"{o}ttlicher, 
C.\,G\"{u}nter,
B.\,Hermberg, 
S.\,Karstensen, 
F.\,Krivan,
A.\,-I.\,Lucaci-Timoce\footnotemark[3], 
S.\,Lu, 
B.\,Lutz, 
S.\,Morozov, 
V.\,Morgunov\footnote{On leave from ITEP}, 
M.\,Reinecke, 
F.\,Sefkow, 
P.\,Smirnov,
M.\,Terwort,
A.\,Vargas-Trevino 
\\ \it
DESY, Notkestrasse 85,
D-22603 Hamburg, Germany
}

\author{\centering  
N.\,Feege, 
E.\,Garutti,
I.\,Marchesini\footnote{Also at DESY}, 
M.\,Ramilli
\\ \it
Univ. Hamburg,
Physics Department,
Institut f\"ur Experimentalphysik,
Luruper Chaussee 149,
22761 Hamburg, Germany
}

\author{\centering 
P.\,Eckert,
T.\,Harion, 
A.\,Kaplan,
 H.\,-Ch.\,Schultz-Coulon,
 W.\,Shen,
 R.\,Stamen,
 A.\,Tadday
\\ \it
 University of Heidelberg, Fakultat fur Physik und Astronomie,
Albert Uberle Str. 3-5 2.OG Ost,
D-69120 Heidelberg, Germany
}

\author{\centering 
B.\,Bilki, E.\,Norbeck, 
Y.\,Onel
\\ \it
University of Iowa, Dept. of Physics and Astronomy,
203 Van Allen Hall, Iowa City, IA 52242-1479, USA
}

\author{\centering 
G.\,W.\,Wilson
\\ \it
University of Kansas, Department of Physics and Astronomy,
Malott Hall, 1251 Wescoe Hall Drive, Lawrence, KS 66045-7582, USA
}

\author{\centering 
K.\,Kawagoe 
\\ \it
Department of Physics, Kyushu University, Fukuoka 812-8581, Japan
}

\author{\centering 
P.\,D.\,Dauncey,  
A.\,-M.\,Magnan
\\ \it
Imperial College London, Blackett Laboratory,
Department of Physics,
Prince Consort Road,
London SW7 2AZ, UK 
}

\author{\centering 
M.\,Wing
\\ \it
Department of Physics and Astronomy, University College London,
Gower Street,
London WC1E 6BT, UK
}

\author{\centering 
F.\,Salvatore\footnote{Now at University of Sussex, Physics and Astronomy Department, Brighton, Sussex, BN1 9QH, UK}
\\ \it
Royal Holloway University of London,
Dept. of Physics,
Egham, Surrey TW20 0EX, UK
}

\author{\centering 
E.\,Calvo~Alamillo, 
M.-C.\, Fouz, 
J.\,Puerta-Pelayo 
\\ \it
CIEMAT, Centro de Investigaciones Energeticas, Medioambientales y Tecnologicas, Madrid, Spain 
}

\author{\centering 
V.\,Balagura, 
B.\,Bobchenko, 
M.\,Chadeeva, 
M.\,Danilov, 
A.\,Epifantsev, 
O.\,Markin,
R.\,Mizuk, 
E.\,Novikov, 
V.\,Rusinov, 
E.\,Tarkovsky
\\ \it
Institute of Theoretical and Experimental Physics, B. Cheremushkinskaya ul. 25,
RU-117218 Moscow, Russia
}

\author{\centering 
N.\,Kirikova,
V.\,Kozlov, 
P.\,Smirnov, 
Y.\,Soloviev 
\\ \it
P.\,N.\, Lebedev Physical Institute,
Russian Academy of Sciences,
117924 GSP-1 Moscow, B-333, Russia
}

\author{\centering 
P.\,Buzhan, B.\,Dolgoshein, A.\,Ilyin, V.\,Kantserov, V.\,Kaplin, A.\,Karakash, E.\,Popova, S.\,Smirnov 
\\ \it
Moscow Physical Engineering Inst., MEPhI,
Dept. of Physics,
31, Kashirskoye shosse,
115409 Moscow, Russia
}

\author{\centering 
C.\,Kiesling,
S.\,Pfau, 
K.\,Seidel, 
F.\,Simon$^\spadesuit$,
C.\,Soldner, 
M.\,Szalay, 
M.\,Tesar,
L.\,Weuste
\\ \it
Max Planck Inst. f\"ur Physik,
F\"ohringer Ring 6,
D-80805 Munich, Germany
}

\author{\centering 
J.\,Bonis, 
B.\,Bouquet,    
S.\,Callier, 
P.\,Cornebise, 
Ph.\,Doublet,
F.\,Dulucq, 
M.\,Faucci Giannelli, 
J.\,Fleury,
H.\,Li\footnote{Now at LPSC Grenoble},  
G.\,Martin-Chassard, 
F.\,Richard, 
Ch.\,de la Taille, 
R.\,P\"{o}schl, 
L.\,Raux,  
N.\,Seguin-Moreau, 
F.\,Wicek
\\ \it

Laboratoire de l'Acc\'{e}l\'{e}rateur Lin\'{e}aire, Centre
Scientifique d'Orsay, Universit\'{e} de Paris-Sud XI, CNRS/IN2P3, BP
34, B\^atiment 200, F-91898 Orsay CEDEX, France
}

\author{\centering 
M.\,Anduze,
V.\,Boudry, 
J-C.\,Brient, 
D.\,Jeans, 
P.\,Mora de Freitas, 
G.\,Musat, 
M.\,Reinhard, 
M.\,Ruan,  
H.\,Videau
\\ \it
 Laboratoire Leprince-Ringuet (LLR)  -- \'{E}cole Polytechnique, CNRS/IN2P3, F-91128 Palaiseau, France
}

\author{\centering 
B.\,Bulanek,
J.\,Zacek 
\\ \it
Charles University, Institute of Particle \& Nuclear Physics,
V Holesovickach 2,
CZ-18000 Prague 8, Czech Republic  
}

\author{\centering 
J.\,Cvach, 
P.\,Gallus, 
M.\,Havranek, 
M.\,Janata, 
J.\,Kvasnicka,
D.\,Lednicky,
M.\,Marcisovsky, 
I.\,Polak, 
J.\,Popule, 
L.\,Tomasek, 
M.\,Tomasek, 
P.\,Ruzicka, 
P.\,Sicho, 
J.\,Smolik, 
V.\,Vrba, 
J.\,Zalesak 
\\ \it
Institute of Physics, Academy of Sciences of the Czech Republic, Na Slovance 2,
CZ-18221 Prague 8, Czech Republic
}

\author{\centering 
B.\,Belhorma,
H.\,Ghazlane
\\ \it
Centre National de l'Energie, des Sciences et des Techniques Nucl\'{e}aires, 
B.P. 1382, R.P. 10001, Rabat, Morocco
}

\author{\centering              
T.\,Takeshita,
S.\,Uozumi
\\ \it
Shinshu Univ.\,,
Dept. of Physics,
3-1-1 Asaki,
Matsumoto-shi, Nagano 390-861,
Japan \\
}

\author{{\centering 
J.\,Sauer, 
S.\,Weber,
C.\,Zeitnitz
\\ \it
Bergische Universit\"{a}t Wuppertal
Fachbereich 8 Physik,
Gaussstrasse 20,
D-42097 Wuppertal, Germany\\
}

\it
$^\spadesuit$ Corresponding author\newline
E-mail: \email{fsimon@mpp.mpg.de}

}

\abstract{The energy resolution of a highly granular 1\,m$^3$ analogue scintillator-steel hadronic calorimeter is studied using charged pions with energies from 10\,GeV to 80\,GeV at the CERN SPS. The energy resolution for single hadrons is determined to be approximately $58\%/\sqrt{E/\mathrm{GeV}}$. This resolution is improved to approximately $45\%/\sqrt{E/\mathrm{GeV}}$ with software compensation techniques. These techniques take advantage of the event-by-event information about the substructure of hadronic showers which is provided by the imaging capabilities of the calorimeter. The energy reconstruction is improved either with corrections based on the local energy density or by applying a single correction factor to the event energy sum derived from a global measure of the shower energy density. The application of the compensation algorithms to {\sc geant}4 simulations yield resolution improvements comparable to those observed for real data.}


\begin{document}


\section{Introduction}

The physics goals of future high-energy lepton colliders such as the ILC \cite{:2007sg} or CLIC \cite{Lebrun:2012hj} put stringent requirements on the detector systems. For example, the efficient event-by-event separation of heavy bosons in hadronic final states requires a jet energy resolution of better than \mbox{4\% \cite{:2007sg}}. This is achievable with Particle Flow Algorithms (PFA) combined with highly granular \mbox{calorimeters \cite{Brient:2002gh,Morgunov:2002pe,Thomson:2009rp}}. The CALICE collaboration has constructed and extensively studied highly granular electromagnetic and hadronic calorimeter prototypes to evaluate detector technologies for future linear collider experiments. These calorimeters have been successfully operated in various test beam experiments in different configurations at DESY, CERN and Fermilab from 2006 until 2012. 
The  unprecedented granularity of the CALICE calorimeter prototypes allows  the structure of hadronic showers to be studied with high spatial resolution, in order to validate different simulation models (for one example of such studies see \cite{Adloff:2010xj}) and to test particle flow algorithms, as demonstrated in \cite{Collaboration:2011ha}. The high granularity also offers the possibility for advanced energy reconstruction methods, the subject of this paper.

We present a study of the hadronic energy resolution of the  CALICE analogue scintillator-steel hadronic calorimeter (AHCAL) \cite{collaboration:2010hb} using data taken at the CERN SPS in 2007 with positive and negative pion beams in the energy range from 10 to 80~GeV.  Two software compensation techniques, which weight energy depositions based on information about the local energy density within the shower obtained from the highly granular readout, are discussed in detail. Both techniques achieve an improvement of the hadronic energy resolution by approximately 20\% for single hadrons in the energy range from 10 to 80~GeV, with a reduction of the stochastic term from  $\sim 58\%/\sqrt{E/\mathrm{GeV}}$ to  $\sim45\%/\sqrt{E/\mathrm{GeV}}$. 

In Section \ref{reco} we briefly describe the test beam setup, discuss the event selection and describe the energy reconstruction, calibration and the determination of the energy resolution in the AHCAL. The software compensation techniques are presented in Section \ref{sc}, and Section \ref{sc:data} summarises the results obtained from data and compares them to simulations.

\section{Energy reconstruction in the AHCAL}
\label{reco}

\subsection{Test beam setup}
\label{reco:setup}
                                                                                   
\begin{figure}
 \begin{center}
  \includegraphics[width=\textwidth]{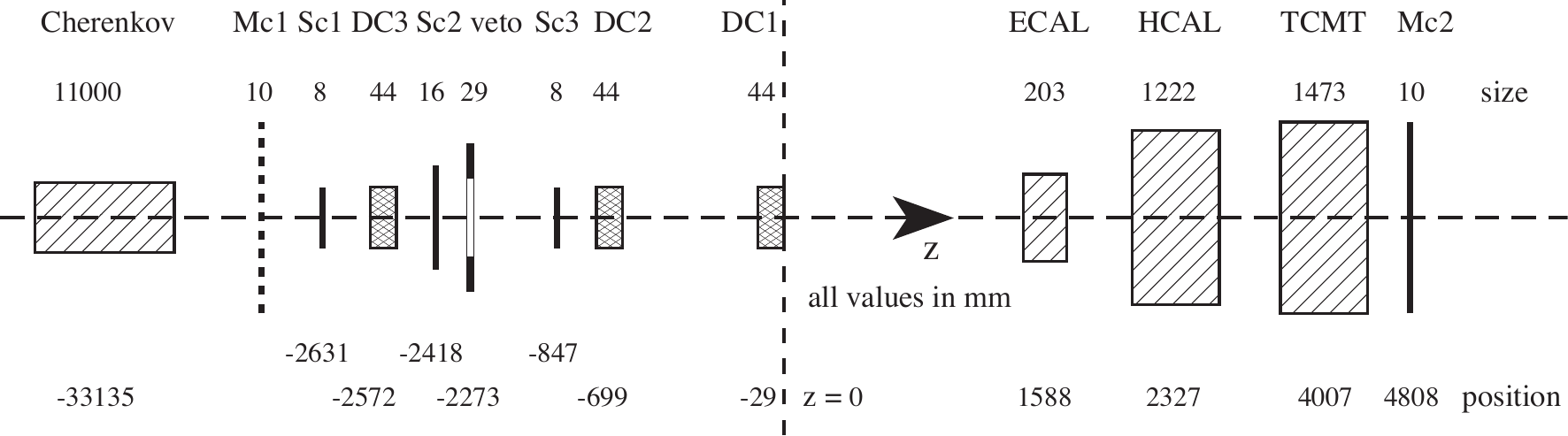}\\
  \caption{Top view of the CALICE test beam apparatus in the CERN SPS H6 beam line including calorimeters, trigger components (scintillator triggers SC1, SC2, and SC3; large area muon trigger counters Mc1, which was installed only during calibration runs, and Mc2; and the beam halo veto), and the tracking drift chambers DC1, DC2, and DC3.  The beam enters from the left.  Dimensions are in millimetres.  Figure is not to scale.  Positions are given at detector centre.}
  \label{fig:setup} 
 \end{center}
\end{figure}

The complete CALICE setup in the H6 beam line at the CERN SPS for the 2007 beam period, illustrated in Figure \ref{fig:setup}, consisted of a silicon-tungsten electromagnetic sampling calorimeter (ECAL)~\cite{Anduze:2008hq}, the AHCAL, and a scintillator-steel tail catcher and muon tracker (TCMT)~\cite{:2012kx}. The test beam setup was also equipped with various trigger and beam monitoring devices.

The ECAL \cite{Anduze:2008hq} has a total depth of 24 radiation lengths (approximately 1 nuclear interaction length $\lambda_{I}$) and consists of 30 active silicon layers arranged in three longitudinal sections with different absorber thicknesses. In this study, the ECAL was used for event selection and early shower detection. Since the present study focuses on the AHCAL, events with a primary inelastic interaction in the ECAL are rejected, as discussed below.

The AHCAL  \cite{collaboration:2010hb} consists of small 5\,mm thick plastic scintillator tiles with embedded wave\-length-shifting fiber and individual readout by silicon photomultipliers (SiPMs). The tiles are assembled in 38 layers with lateral dimensions of 900$\times$900 mm$^2$, separated by 21 mm of steel. The absorber material in each layer is made of 17~mm thick absorber plates and two 2 mm thick cover plates of the cassettes that house the scintillator cells. The size of the scintillator tiles ranges from \mbox{$30\times 30\; \textrm{mm}^2$} in the central region and  \mbox{$60\times 60\; \textrm{mm}^2$} in the outer region to \mbox{$120\times 120\; \textrm{mm}^2$} along the perimeter of each layer. In the last eight layers only \mbox{$60\times 60\; \textrm{mm}^2$} and \mbox{$120\times 120\; \textrm{mm}^2$} tiles  are used. In total, the CALICE AHCAL has 7608 scintillator cells and a thickness of 5.3 $\lambda_{I}$ (4.3\,$\lambda_{\pi}$). To be able to correct off-line for variations in response of the photon sensors caused by the substantial temperature variations in the experimental hall, the temperature of the AHCAL inside the readout cassettes is monitored in each layer by five sensors installed equally spaced from bottom to top in the centre of the detector. 

The TCMT \cite{:2012kx} consists of 16 readout layers assembled from 5 mm thick,  50 mm wide and \mbox{1000 mm} long scintillator strips with embedded wavelength-shifting fibers read out at one end by SiPMs. Each of the scintillator layers has 20 strips for a total of 320 strips in the TCMT. The scintillator is sandwiched between steel absorber plates. The TCMT has two sections with different sampling fractions, one fine section with 21 mm thick absorbers for the first nine layers (where the absorber plate for the first layer is the back plate of the AHCAL), and a coarse section with 104 mm thick absorbers. Two mm of the absorber thickness in each layer is provided by the cover sheets of the scintillator strip cassettes. In this study the information from the TCMT is used for muon separation and to measure energy leaking out the back of  the AHCAL, which is of  particular importance at higher energies. The total depth of the CALICE calorimeter setup amounts to approximately 12\,$\lambda_{I}$, with a total of 17\,648 readout channels. 

In addition to the calorimeters themselves, the setup includes auxiliary detectors for triggering, tracking and particle identification as shown in Figure \ref{fig:setup}. The scintillation counters Sc1, Sc2 and Sc3 provide the beam trigger, where a coincidence between at least two out of the three is required. In addition, Sc2 has an analogue readout to tag multi-particle events. The large area veto counter is used to reject beam halo events and a large area scintillator counter Mc2 downstream of the TCMT provides muon tagging for particles penetrating the full calorimeter setup. For dedicated muon runs, an additional large area scintillation counter, Mc1, is installed upstream of the calorimeters. Three drift chambers DC1, DC2 and DC3 determine the position of the incoming beam particles. Particle identification is provided by a threshold  \u{C}erenkov counter upstream of the calorimeters, which discriminates between electrons and pions or between pions and protons in negatively or positively charged beams, respectively, by appropriately chosen gas pressures.

\subsection{Event selection}
\label{reco:selection}

The response of the individual calorimeter cells is calibrated with muons, using the visible signal of a minimum-ionising particle (MIP) as the cell-to-cell calibration scale. This signal corresponds to 13 detected photo-electrons in typical cells. After this cell-to-cell calibration, the most probable energy loss of a MIP is used as the base unit of the energy measurement. To reject noise, only cells with a visible energy above a threshold of 0.5~MIP are used in the analysis, referred to as hits in the following. 

\begin{table}
\centering
\begin{tabular}{c|c|r|r}
particle type & beam energy  [GeV]  &  all pions  &  selected pions \\
\hline
\hline
 $\pi^{-}$  &  10    &  440208  &  84706 \\
 $\pi^{-}$  &  15    &  127554  &  24997 \\
 $\pi^{-}$  &  18   &   52880  &  10492\\
 $\pi^{-}$  &  20    &  342798  &  67093\\
 $\pi^{-}$  &  25    &  201243  &  39631\\
 $\pi^{-}$  &  35    &  272987  &  54126\\
 $\pi^{-}$  &  40    &  472345  &  93301\\
 $\pi^{-}$  &  45    &  325092  &  63547\\
 $\pi^{-}$  &  50    &  304023  &  59076\\
 $\pi^{-}$  &  60    &  647090  & 121588\\
 $\pi^{-}$  &  80    &  741440  & 139248 \\
 \hline
 $\pi^{+}$  &  30    &  155210  &  30884\\
 $\pi^{+}$  &  40    &  307177  &  60595\\
 $\pi^{+}$  &  50    &  159414  &  30843\\
 $\pi^{+}$  &  60    &  449273  &  86947\\
 $\pi^{+}$  &  80    &  272441  &  52442 \\
\end{tabular}

\caption{Summary of the data samples. The total number of pions is the number of events classified as pions, after rejection of empty, noisy and double particle events, and the application of muon rejection and particle identification cuts. The number of selected pions are the events with an identified shower start in the first five layers of the AHCAL, which are used in the present analysis. For most energies, several run periods at different temperatures are combined to maximise statistics. \label{tab:EventStatistics}}
\end{table}

The data samples for the present analysis are selected from $\pi^{-}$ and $\pi^{+}$  data in the energy range of 10 to 80\,GeV and 30 to 80\,GeV respectively, as summarised in Table \ref{tab:EventStatistics}. To maximise statistics, data from several run periods taken at different temperatures are combined for most energies, with corrections for the temperature dependence of the response of the photon sensors applied during event reconstruction \cite{Thesis:Feege}. The event selection procedure purifies the pion sample by rejecting admixtures of muons, electrons, and protons. To identify muons, information from the ECAL, AHCAL and TCMT is used, requiring small energy deposits consistent with a minimum-ionising particle in all three detectors. Optimal separation of muons and hadrons is achieved by using beam energy-dependent constraints on the energy sum in the TCMT versus the combined energy sum of the ECAL and AHCAL. For beam energies of 30\,GeV and 35\,GeV a muon contamination at the level of 30\% and 15\% before muon rejection is observed, respectively. For all other energies the muon content does not exceed 7\%. After the event selection, the muon content is below 0.5\% at all energies, estimated using the muon identification efficiency of 98\% at 10 GeV and 99.5\% at 30 GeV and above, which is determined from muon data and simulations. Protons and kaons are removed from the  $\pi^{+}$ samples by requiring a positive pion identification in the \u{C}erenkov counter. Even before selection, the kaon content is below 3\% at all energies. The proton content of the beam is very small below 30 GeV since a tertiary beam is used at these energies, and varies between approximately 15\% and 30\% at higher energies. Since the positive identification of $\pi^{+}$ is based on the detection of  \u{C}herenkov photons the proton and kaon contamination of the positive pion sample is negligible. Electrons are removed from the $\pi^{-}$ sample both by the  \u{C}erenkov counter and by selecting events with no inelastic interaction in the ECAL, as discussed below. 

Since the goal of the present analysis is the study of the performance of the AHCAL, pion showers that develop predominantly in the AHCAL are selected. This is achieved by requiring that the position of the primary inelastic interaction is located in the first five layers of the hadron calorimeter. This excludes events with sizeable energy deposit in the ECAL while keeping energy leakage into the TCMT to a minimum. The location of the primary inelastic interaction is determined by detecting the change from a minimum-ionising particle track to multiple secondary particles, evidenced by increased energy deposition and number of hits over several consecutive layers \cite{Collaboration:2011ha}. Simulation studies indicate that the difference between the reconstructed and the true primary interaction layer does not exceed one layer for 78\% of all events and does not exceed two layers for more than 90\% of all events in the energy range from 10 to 80~GeV.

\subsection{Energy reconstruction  and intrinsic energy resolution}
\label{reco:reco}

To measure the energy deposited in the sub-detectors, a conversion from the visible signal in MIP units to the total energy in units of GeV is necessary. Since only hadrons with a shower start in the AHCAL are considered, the relevant conversion factor for the ECAL is determined using simulated muons to obtain the correlation between the visible energy and the true ionisation energy loss in the detector. This factor is validated with the  measured response to muons obtained from a sample of muon data. The ratio of the visible signal in the active silicon to the total deposited energy in active and passive material  is approximately 25\% higher for minimum-ionising particles than for electromagnetic showers, resulting in a lower conversion factor than that for electrons presented in \cite{Adloff:2009zz}. The total energy deposited in the AHCAL is obtained at the electromagnetic scale, using calibration factors determined for electron and positron data \cite{collaboration:2010rq}.  Since the AHCAL is a non-compensating calorimeter, the response to hadrons differs from that to electrons, requiring an additional scaling factor. This factor is determined by comparing the reconstructed energy for pions using the electromagnetic calibration factors with the known beam energy. In the present study, the energy dependence of this factor is ignored by taking a constant $\frac{e}{\pi} = 1.19$, corresponding to the average over the energy range studied. Since the first nine TCMT layers are essentially identical to the AHCAL layers in terms of absorber and active material, the same electromagnetic calibration factors and an identical $\frac{e}{\pi}$ ratio are assumed. For the last seven TCMT layers, the calibration factors are adjusted according to the increased absorber thickness.

For each event, the uncorrected reconstructed energy for hadrons, $E_{\mathrm{unc}}$, is given by the sum of reconstructed energies in the three calorimeters, 
\begin{equation}
E_{\mathrm{unc}} = E_{\mathrm{ECAL}}^{\mathrm{track}} \,+\, \frac{e}{\pi} \cdot  \left(E_{\mathrm{HCAL}} \,+\, E_{\mathrm{TCMT}}\right),
\label{eq:eventEini}
\end{equation}
where $E_{\mathrm{ECAL}}^{\mathrm{track}}$ is the measured energy in the ECAL deposited by the particle track, and $E_{\mathrm{HCAL}}$  and $E_{\mathrm{TCMT}}$ are the energies measured in the AHCAL and in the TCMT, both given at the electromagnetic scale. The energy in each subdetector is given by the sum of all hits above a noise threshold of 0.5 MIP.

The resulting reconstructed energy distributions are fit with a Gaussian in the interval of $\pm 2$ standard deviations around the mean value, providing good fits for all energies. The differences compared to a fit over the full range are on the sub-percent level for the extracted mean and on the one percent level for the standard deviation and depend on the beam energy. Fitting over the full range reduces the fit quality for some energies in particular for the uncorrected data, leading to the choice of $\pm 2$ standard deviations for best consistency between the different data points. In the following, the mean and standard deviation of this Gaussian fit at a given beam energy are referred to as the mean reconstructed energy $E_{\mathrm{reco}}$ and the resolution $\sigma_{\mathrm{reco}}$, respectively. 

Systematic uncertainties on the energy measurement in the AHCAL are discussed in detail in \cite{collaboration:2010rq}. For the reconstruction of hadrons, the main uncertainty is due to the MIP to GeV conversion factor that is extracted from the electromagnetic calibration of the detector. The size of the uncertainty was studied thoroughly for the present data set, and is determined to be 0.9\% by varying the calibration constants within the allowed limits. Other effects which contribute to the uncertainties for electromagnetic showers, such as the saturation behaviour of the photon sensor, are found to be negligible for hadrons even at the highest energies studied here. 

\begin{figure}
 \begin{center}
  \includegraphics[width=0.49\textwidth]{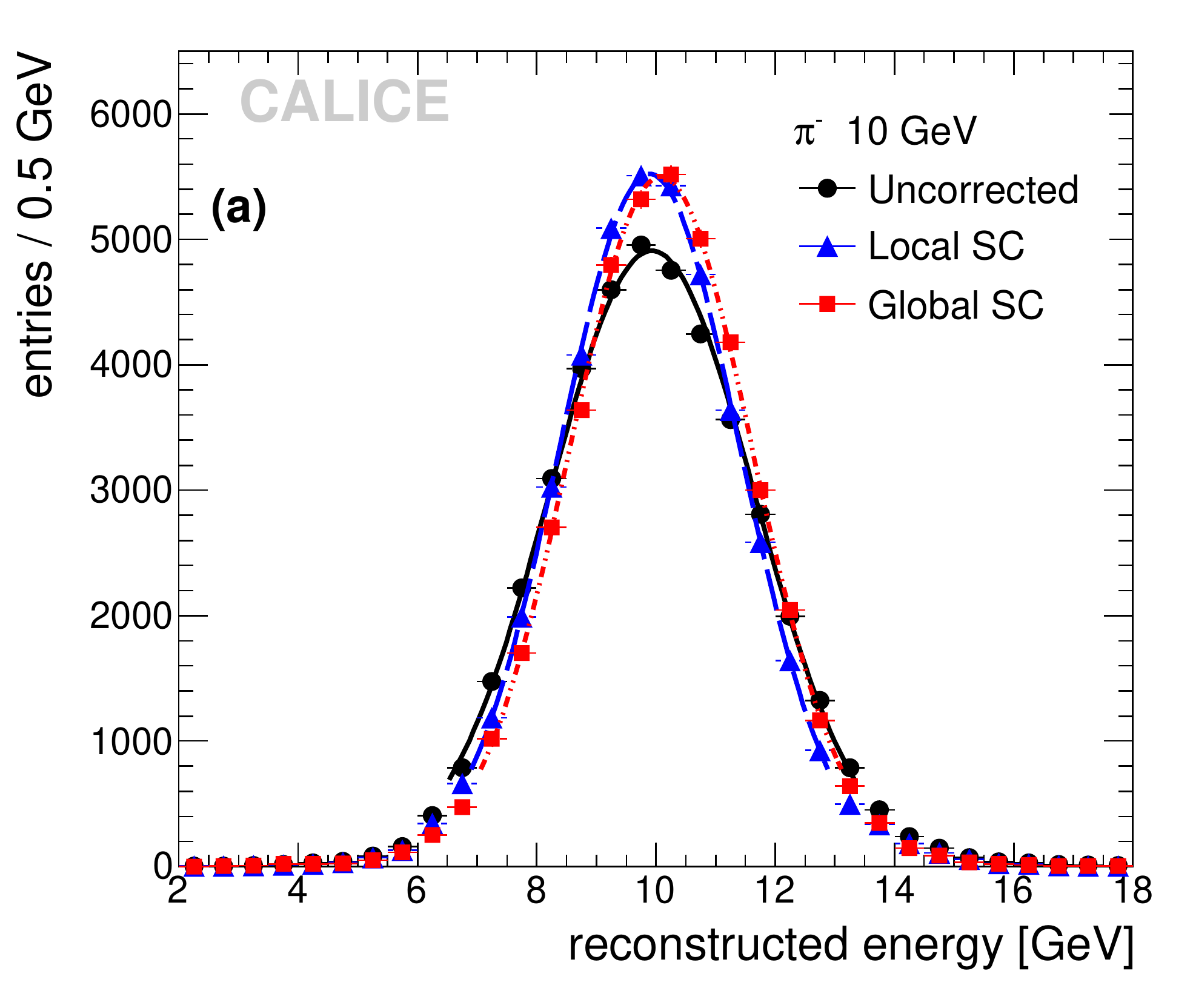}
 \hfill
  \includegraphics[width=0.49\textwidth]{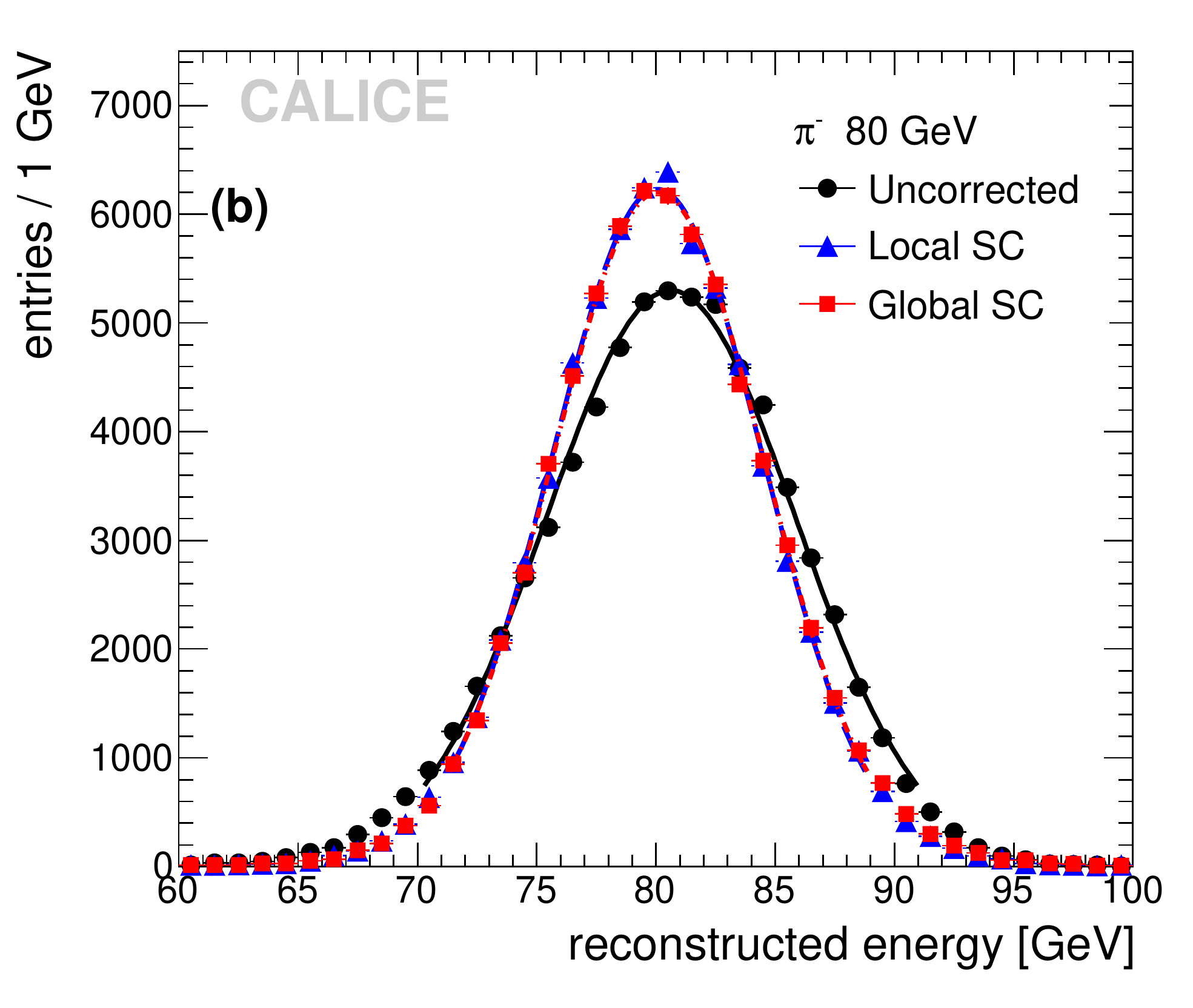}
  \caption{Reconstructed energy distributions for  10~GeV $\pi^-$ (a) and 80~GeV $\pi^-$ (b) without compensation (black circles) and after local software compensation (LC), shown by the blue triangles, and after global software compensation (GC), shown by the red squares. The curves show Gaussian fits to the distributions in the range of $\pm 2$ standard deviations. Errors are statistical only.}
  \label{fig:enr_dist}
 \end{center}
\end{figure}

Figure \ref{fig:enr_dist} shows the distribution of reconstructed energies for 10 GeV and 80 GeV pions, with the uncorrected reconstructed energy shown by black data points. At all energies, the distributions of the reconstructed energies follow a Gaussian distribution well, with typically more than 95\% of all events in the fit range of $\pm 2$ standard deviations. The software compensation methods also included in the figure are described in Sections \ref{sc:local} (local software compensation) and \ref{sc:global} (global software compensation).

\begin{figure}
 \begin{center}
  \includegraphics[width=0.7\textwidth]{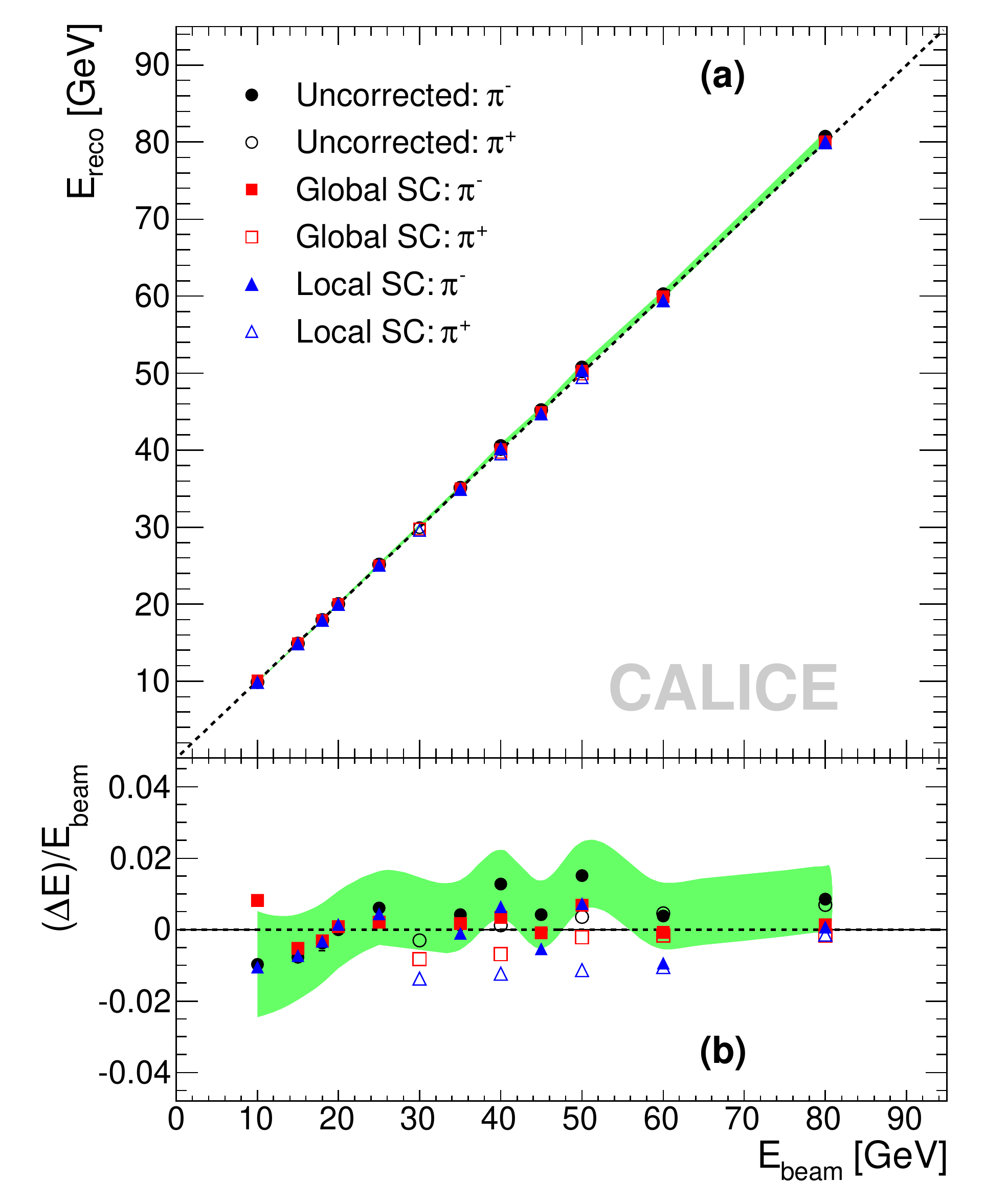}\\
\caption{(a) Mean reconstructed energy for pions and (b) relative residuals to beam energy versus beam energy without compensation (black circles) and after local software compensation (LC), shown by the blue triangles, and after global software compensation (GC), shown by the red squares. Filled and open markers indicate $\pi^{-}$ and $\pi^{+}$, respectively. Dotted lines correspond to $E_{\mathrm{reco}} = E_{\mathrm{beam}}$. Systematic uncertainties are indicated by the green band, which corresponds to the uncertainties for the uncorrected $\pi^{-}$ data sample.}
  \label{fig:lin_data}
 \end{center}
\end{figure}

Figure \ref{fig:lin_data} shows the mean reconstructed energy versus beam energy, with the black points giving the uncorrected reconstructed energy. The measured responses to positive and negative pions agree well within the systematic uncertainties, which are shown by the green band. Relative residuals to the beam energy are shown in the lower panel of Figure \ref{fig:lin_data}, where $\Delta E = E_{\mathrm{reco}} - E_{\mathrm{beam}} $. The linearity of the calorimeter response to hadrons showering predominantly in the AHCAL is within $\pm$2\% in the studied energy range.

\begin{figure}
 \begin{center}
  \includegraphics[width=0.7\textwidth]{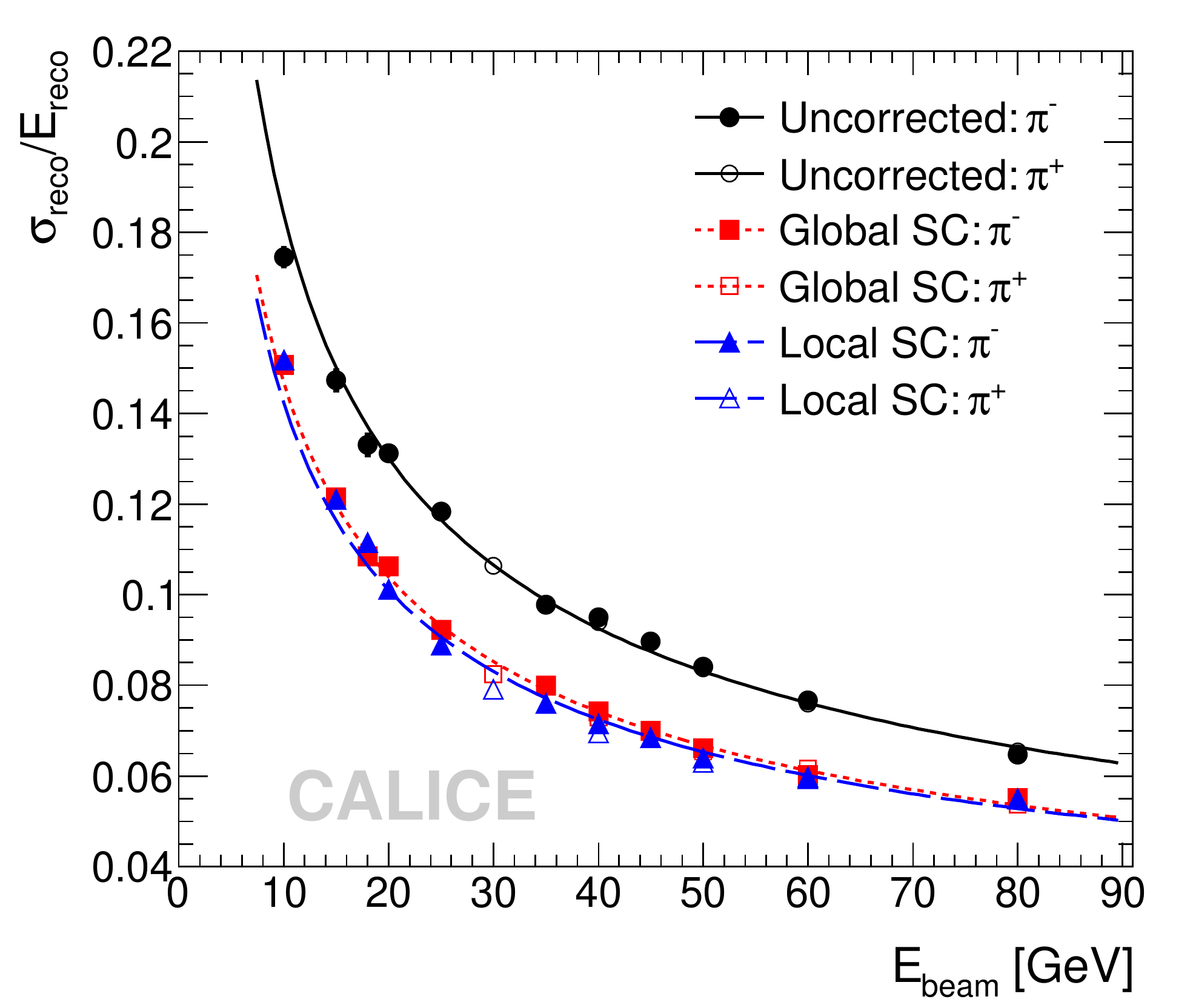}
  \caption{Energy resolution versus beam energy without compensation and after local and global software compensation. The curves show fits using Equation {\protect \ref{eq:relres}}, with the black solid line showing the fit to the uncorrected resolution, the red dotted line to the global software compensation and the blue dashed line to the local software compensation. The stochastic term is $(57.6\pm0.4)\%$, $(45.8\pm0.3)\%$ and $(44.3\pm0.3)\%$, with constant terms of $(1.6\pm0.3)\%$, $(1.6\pm0.2)\%$ and $(1.8\pm0.3)\%$ for the uncorrected resolution, global software compensation and local software compensation, respectively.}
  \label{fig:res_data}
 \end{center}
\end{figure}

The fractional energy resolution, $\sigma_{\mathrm{reco}}/E_{\mathrm{reco}}$, is shown in Figure \ref{fig:res_data}. Again, the uncorrected resolution is indicated by black points. The measured resolution for $\pi^{-}$ is in very good agreement with that obtained for $\pi^{+}$, with the differences smaller than the size of the markers for all energies where both $\pi^{-}$ and $\pi^{+}$ results exist. The black solid curve shows the result of a fit to these points with the following function:
\begin{equation}
 \frac{\sigma_{\mathrm{reco}}}{E_{\mathrm{reco}}} = \frac{a}{\sqrt{E_{\mathrm{beam}}}} \oplus b \oplus \frac{c}{E_{\mathrm{beam}}},
 \label{eq:relres}
\end{equation}
where $E_{\mathrm{beam}}$ is the beam energy in GeV, and $a$, $b$ and  $c$ are the stochastic, constant and noise contributions, respectively. The noise term is fixed to $c = 0.18$~GeV, corresponding to the measured noise contribution in the full CALICE setup taking into account contributions from the ECAL (0.004~GeV), the AHCAL (0.06~GeV) and the TCMT (0.17~GeV). These values are obtained from the standard deviation of the noise levels measured in dedicated runs without beam particles as well as in random trigger events constantly recorded during data taking. From the fit, the stochastic term of the uncorrected hadron energy resolution of the AHCAL is determined to be $(57.6 \pm 0.4)\%/\sqrt{E/\mathrm{GeV}}$ and the constant term to be $(1.6 \pm 0.3)\%$.

\section{Software compensation: motivation and techniques}
\label{sc}

In ideal sampling calorimeters the energy measured for electromagnetic showers is directly proportional to the incoming particle energy. In the absence of instrumental effects such as non-linearities or saturation of the readout, the energy of a particle can thus be obtained by multiplying the visible signal by a single energy-independent factor accounting for the non-measured energy depositions in the passive absorber material. 

The calorimeter response to hadron-induced showers is more complicated \cite{Fabjan:1976da}, since these showers have contributions from two different components: an electromagnetic component, originating primarily from the production of $\pi^0$s and $\eta$s and their subsequent decay into photon pairs; and a purely hadronic component. The latter includes ``invisible'' components from the energy loss due to the break-up of absorber nuclei, from low-energy particles absorbed in passive material and from undetected neutrons, depending on the active material. This typically leads to a reduced response of the calorimeter to energy in the hadronic component, and thus overall to a smaller calorimeter response to hadrons compared to electromagnetic particles of the same energy. Since the production of  $\pi^0$s and $\eta$s are statistical processes, the relative size of the two shower components fluctuates from shower to shower, which, combined with the differences in visible signal for electromagnetic and purely hadronic energy deposits, leads to a deterioration of the energy resolution. In addition, the average fraction of energy in the electromagnetic component depends on the number of subsequent inelastic hadronic interactions and thus on the initial particle energy. The electromagnetic fraction of hadronic showers increases with increasing particle energy \cite{Gabriel:1993ai}, often resulting in a non-linear response for  non-compensating calorimeters.  

There are two fundamentally different approaches to improve the energy resolution of a ha\-dron\-ic sampling calorimeter. One approach is to eliminate the issue of different response to electromagnetic and hadronic components by design, through the construction of so-called compensating calorimeters. This can be achieved by specific choices of absorber and active material which enhance the sensitivity to neutrons, and thus to the hadronic component of the shower, and by appropriately chosen sampling fractions. However, these conditions impose very strict requirements on the materials  used and on the overall geometry of the whole detector system. One prominent example of a compensating calorimeter is the uranium-scintillator calorimeter of the ZEUS experiment~\cite{Derrick:1991tq, Andresen:1991ph}, which reached a stochastic resolution term of  $34.5\%/\sqrt{E/\mathrm{GeV}}$ for single pions~\cite{d'Agostini:1988dd}. 

On the other hand, for intrinsically non-compensating calorimeters, compensation can be achieved by so-called ``off-line weighting'' or ``software compensation'' techniques. These techniques assign different weights to electromagnetic and hadronic energy deposits on an event-by-event basis. The differing spatial structure of the electromagnetic and hadronic components of particle showers can be used to characterise the origin of energy deposits. Since the radiation length is much shorter than the nuclear interaction length in heavy absorbers used in hadronic calorimeters, electromagnetic sub-showers are more compact than purely hadronic sub-showers, generally resulting in a higher energy density of the electromagnetic component. The application of software compensation techniques relies on longitudinal and lateral segmentation of the calorimeters, to provide the necessary information for a measurement of the energy density of particle showers. One of the first applications of such techniques was in the WA1/CDHS scintillator steel calorimeter, where an improvement of the hadronic resolution between 10\% and 30\% was achieved in the energy range of 10~GeV to 140~GeV  \cite{Abramowicz:1980iv}. These techniques were further refined and applied in various experiments, such as the H1 liquid argon calorimeter \cite{Andrieu:1993tz} and the ATLAS calorimeter system~\cite{Cojocaru:2004jk}.

With its unprecedented high granularity, the CALICE AHCAL is well suited for such techniques. In the present paper, two techniques based on an event-by-event analysis of the hit energy distributions are discussed. The local software compensation (LC) procedure is based on a re-weighting of each individual hit depending on the local energy density. The global software compensation (GC) procedure uses the distribution of hit energies to derive one global factor for the correction of the reconstructed energy of the complete hadronic shower. The parameters used for both techniques are determined from test beam data, as discussed in detail below. The available data set is split into two samples of equal event count, a training data set and the data set used to study the energy reconstruction. This ensures a statistical independence of the data used to determine the parameters for the software compensation algorithms and the data used to evaluate the performance of the techniques.

\subsection{Local software compensation}
\label{sc:local}

The local software compensation technique improves the energy reconstruction for hadrons by applying weights to the energy recorded in every cell of the AHCAL within a hadronic shower. The weights are chosen based on the local energy density, which is taken as a measure of the likelihood that a given cell belongs to an electromagnetic or a hadronic sub-shower. In the present study, the energy content of a cell, divided by its volume, is taken as the relevant local energy density. Electromagnetic sub-showers typically have a higher energy density than purely hadronic ones, and, due to the non-compensating nature of the AHCAL, result in a larger detector signal per unit of deposited energy. Thus, cells with  a higher energy content are assigned a lower weight in the total energy sum than cells with a low energy content to correct for this difference.  
 The reconstructed energy of each event corrected with local software compensation, $E_{\mathrm{LC}}$, is thus given by introducing weights for each AHCAL hit in Equation \ref{eq:eventEini}, resulting in
\begin{equation}
 \label{eq:local-e-sum}
  E_{\mathrm{LC}} = E_{\mathrm{ECAL}}^{\mathrm{track}} + \frac{e}{\pi} \cdot  \left( \sum_{i} \left( E_{\mathrm{HCAL},i} \cdot \omega_{i} \right) + E_{\mathrm{TCMT}}\right)
\end{equation}
where $\omega_{i}$ is the energy density dependent weight applied to the cell energy $E_{\mathrm{HCAL},i}$.

\begin{figure}
 \centering
  \includegraphics[width=0.5\textwidth]{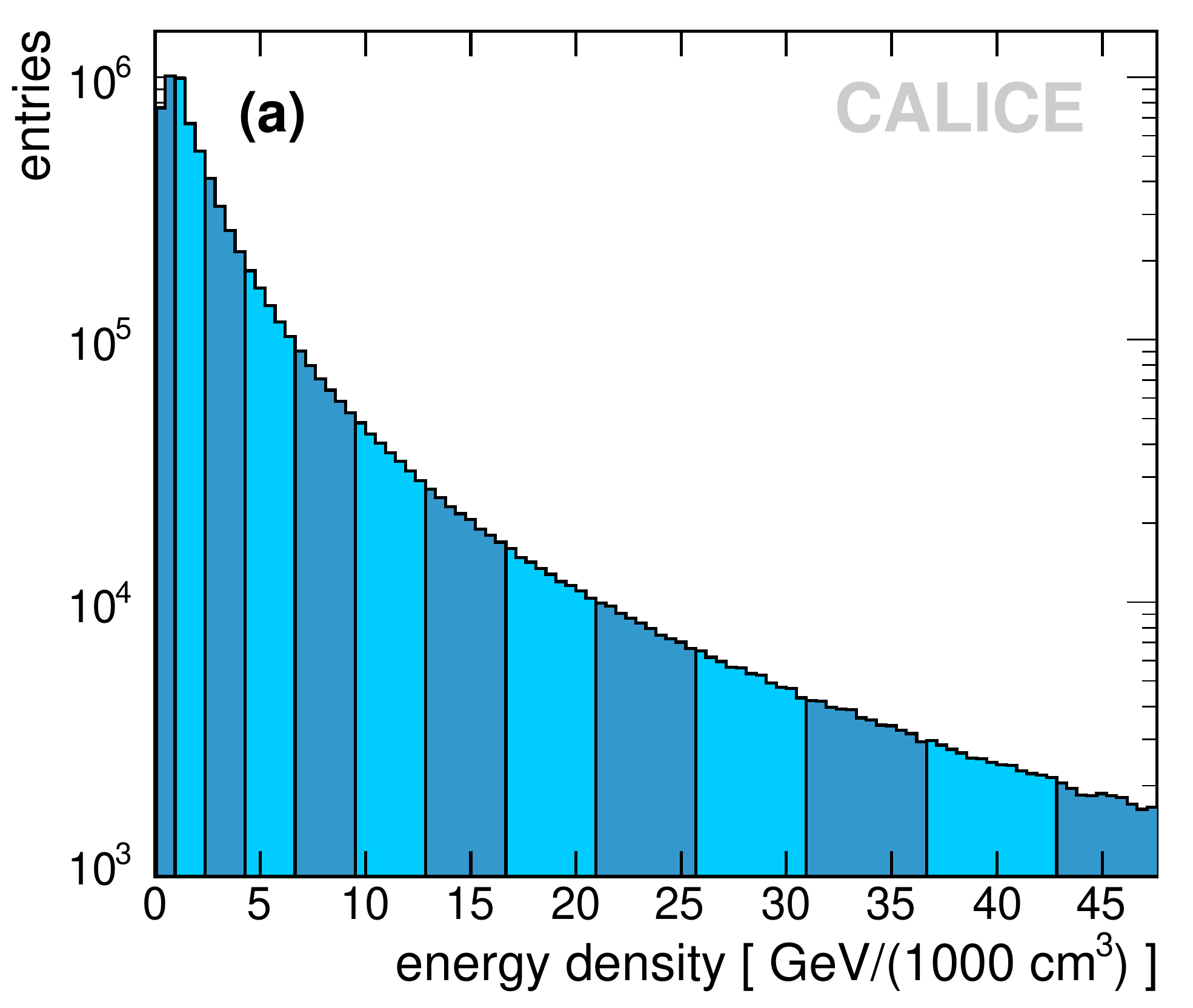}\hfill	
  \includegraphics[width=0.5\textwidth]{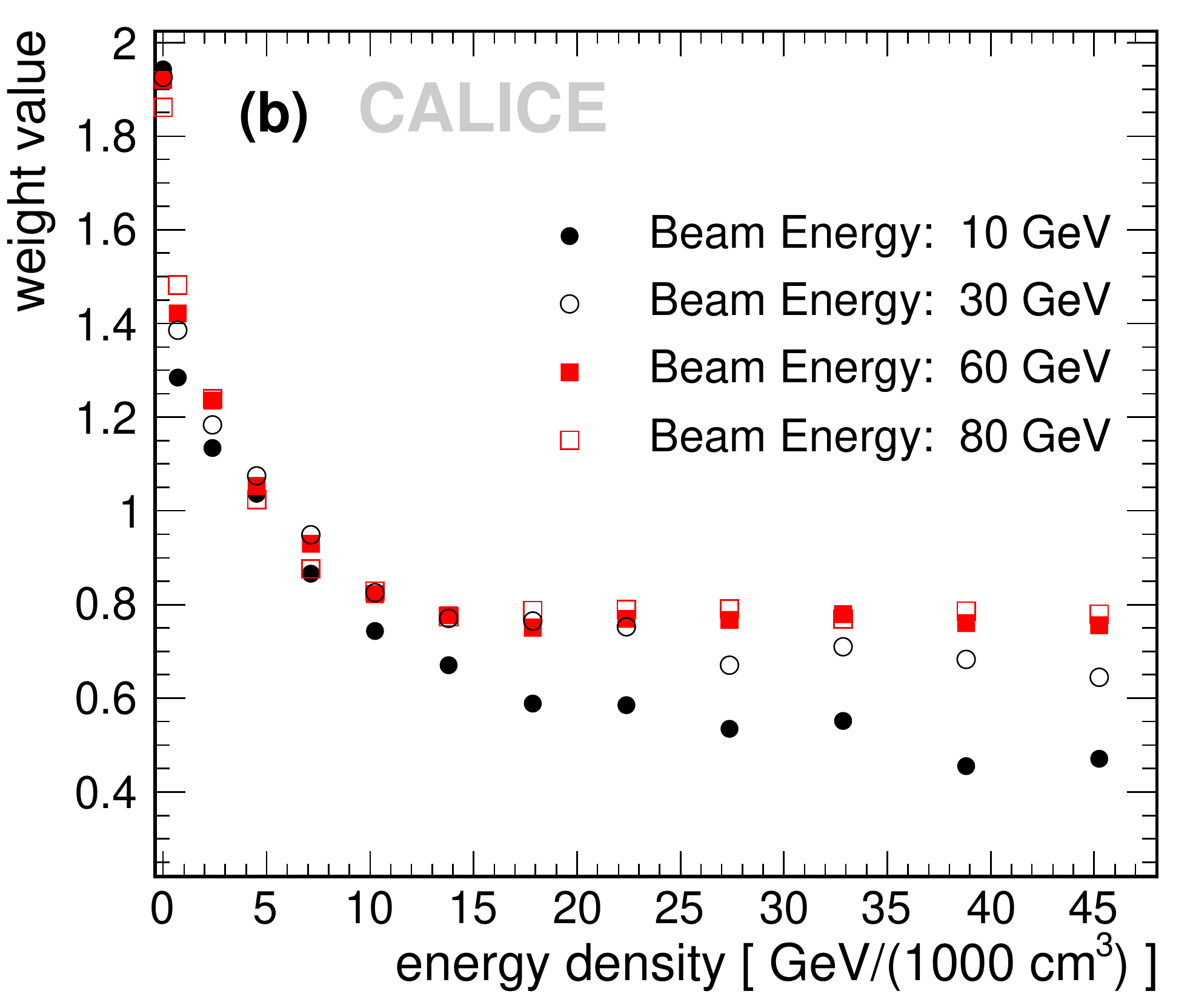}
  \caption{(a) Distribution of the cell energy density in the AHCAL for 40 GeV pion showers. The energy density is given by the uncorrected reconstructed energy in that cell, divided by the corresponding absorber volume. The different energy density bins used in the analysis are indicated by colour shades. (b) Optimal weights as a function of energy density for different beam energies, determined without constraints of a specific functional form in the first iteration of the minimisation.}
  \label{fig:bins_LC}
\end{figure}

To make the technique robust against fluctuations entering due to the relatively low number of hits in a given event, the single cell energy density distribution is subdivided into bins in energy density, as illustrated in Figure \ref{fig:bins_LC} (a).  The binning is also needed for the minimisation technique chosen here for the determination of the weights as discussed below. For each bin, a separate weight is determined which is applied to all hits that fall into that particular bin. The number of sub-divisions in energy density is chosen as a compromise between the requirements for fine subdivisions to maximise the sensitivity of the algorithm to differences in shower structure on one hand, and the stability of the determination of the weights and of the algorithm on the other hand. While a fine binning improves the sensitivity to the shower structure, a robust determination of the weights requires sufficient statistics in each bin, and changes of the weights from bin to bin. The performance of the local software compensation does not depend on the precise choice of bin number and bin borders.

Since the overall energy density of hadronic showers changes with energy, the weights $\omega$ depend both on the cell energy density $\rho$ and on the particle energy. The weights, as a function of energy density and particle energy, are determined from the training data set extending over the full energy range studied here. The optimal weights are found by minimising a simplified $\chi^2$ given by the function $\chi^{2} = \sum_i (E_{\mathrm{LC},i} - E_{\mathrm{beam}})^2$, where $E_{\mathrm{LC},i}$ is the reconstructed energy of a given event using software compensation, and the sum runs over all events used for the weight determination. In this minimisation, the bin by bin weights are used as free parameters. \mbox{Figure \ref{fig:bins_LC} (b)} shows the optimal weights determined with this procedure for four different energies. The weights at a given beam energy can be parametrized by 
\begin{equation}
\label{eq:local-weight-function}
 \omega = p_0 + p_1 \cdot \exp(p_2 \cdot \rho), 
\end{equation}
where $\rho$ is the energy density corresponding to the centre of the energy density bins introduced above, and $p_0, p_1$ and $p_2$ are parameters of the weight function. These parameters depend on the beam energy, with their energy dependence following exponential functions in particle energy for $p_0$ and $p_1$, and a logarithmic function in particle energy for $p_2$. A robust determination of the weights is achieved by an iterative minimisation procedure, where the free parameters $p_0$, $p_1$ and $p_2$ are consecutively fixed to the function determined in the previous minimisation stage.  

For the application of this technique to data, no {\it a priori} knowledge of the particle energy is required, as the uncorrected reconstructed particle energy is used instead of $E_{\mathrm{beam}}$ to select the correct weight parametrisation. Since the energy dependence of the weight parameters is not very steep, this does not introduce a noticeable bias for the reconstructed energy. A second iteration does not lead to significant further improvement and is thus not performed in the reconstruction.

\subsection{Global software compensation}
\label{sc:global}

The global software compensation technique improves the energy resolution for hadrons by applying a single weight to the reconstructed shower energy. This weight is derived from the distribution of hit energies in the hadronic shower, providing sensitivity to the overall energy density, and thus to the fraction of hits in electromagnetic sub-showers. Since electromagnetic sub-showers are characterised by a high local energy density, a hadronic shower with a large electromagnetic content will have a larger fraction of high-energy hits than a shower with predominantly hadronic contributions. 

\begin{figure}
 \begin{center}
  \includegraphics[width=0.7\textwidth]{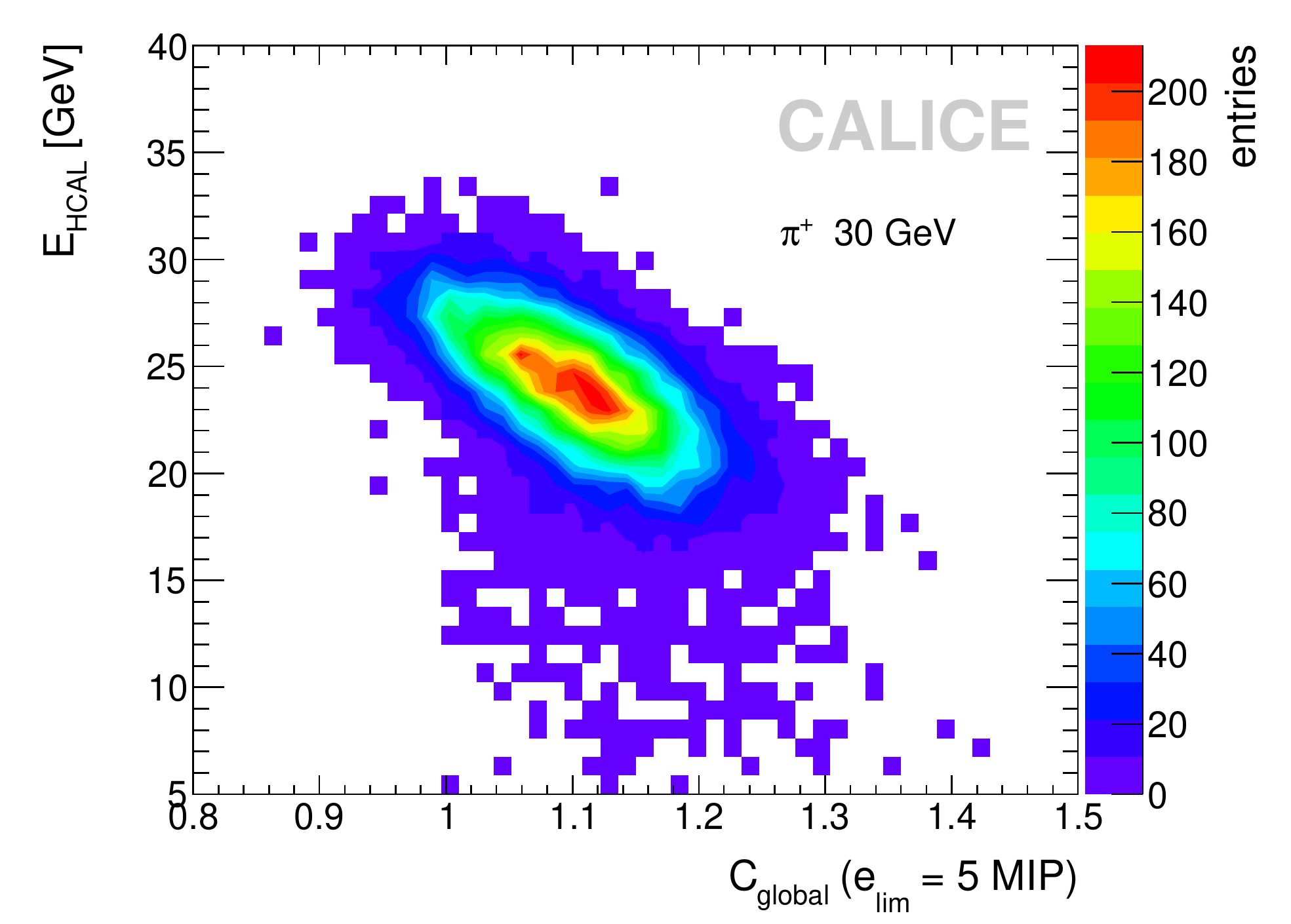}\\
  \caption{Correlation of the factor $C_{\mathrm{global}}$ and the reconstructed energy in the AHCAL, $E_{HCAL}$, for showers induced by $\pi^{+}$ at 30~GeV.}
  \label{fig:gl_corr}
 \end{center}
\end{figure}

The determination of the event weight is based on a phenomenological approach using the fraction of calorimeter hits below a certain energy threshold. This fraction is computed for each event, and provides a measure of the importance in each shower of low-density energy deposits, which are expected to be predominantly of hadronic origin. Based on this, with an additional consideration of the overall hit energy distribution given by the number of hits below the mean energy value of the hit energy, the factor $C_{\mathrm{global}}$ is constructed, which is used to correct the reconstructed energy. This factor, calculated for each event, is given by the ratio of the number of shower hits with a measured visible signal below a given threshold $e_{\mathrm{lim}}$ and the number of shower hits with a measured visible signal below the mean value of the hit energy spectrum for that particular event. Figure \ref{fig:gl_corr} illustrates the sensitivity of the factor $C_{\mathrm{global}}$ to the reconstructed energy, for a value of $e_{\mathrm{lim}}$ = 5 MIP applied to $\pi^{+}$ events at 30\,GeV. The clear anti-correlation between the reconstructed energy and $C_{\mathrm{global}}$ provides the basis for an improved energy reconstruction using this factor. The anti-correlation is due to the fact that events with a high electromagnetic content tend to have a larger number of high-energy hits above $e_{\mathrm{lim}}$  and thus a lower $C_{\mathrm{global}}$, while those events have a higher reconstructed energy.

\begin{figure}
 \begin{center}
  \includegraphics[width=10cm]{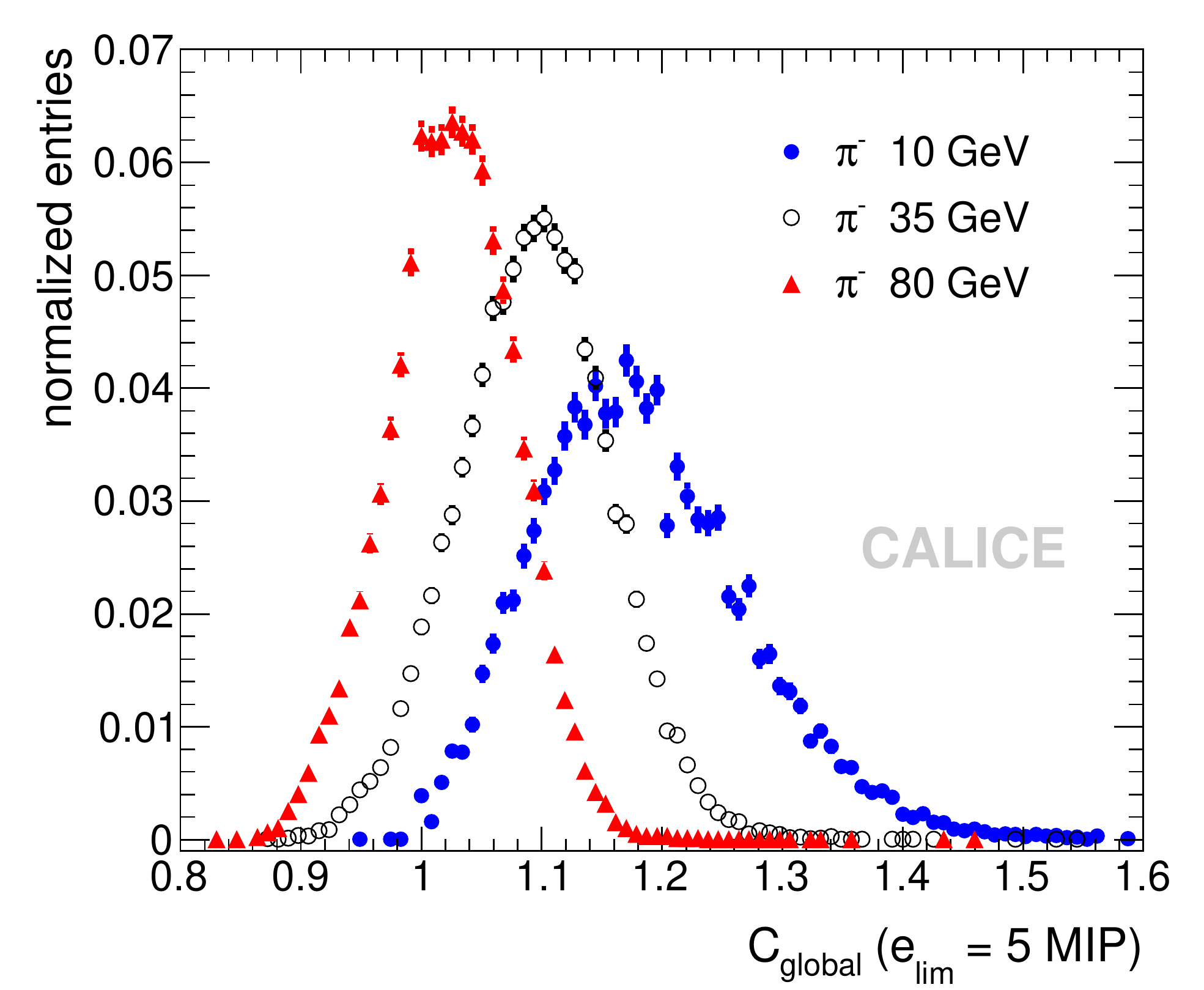}\\
  \caption{Distributions of the factor $C_{\mathrm{global}}$ for hadronic showers induced by  $\pi^{-}$ with an energy of 10 GeV (blue squares), 35 GeV (black circles) and 80~GeV (red triangles), respectively. Statistical errors are shown.}
  \label{fig:gl_compfac}
 \end{center}
\end{figure}

The value of $e_{\mathrm{lim}}$ was optimised to provide good performance of the algorithm over the full energy range, with the linearity of the detector response taken as a key factor. While higher values for $e_{\mathrm{lim}}$ provide stricter separation of electromagnetic and non-electromagnetic events, if the value is set too high this results in asymmetric distributions of $C_{\mathrm{global}}$ at lower energy, leading to reduced performance. These asymmetries originate from the reduced number of high-energy hits at low particle energies. For example, a large fraction of 10 GeV pion showers have essentially no hits above 7 MIP.  Too low values, on the other hand, result in a non-linear response due to the reduced sensitivity to the electromagnetic component at higher particle energies. Best performance was obtained for a value of $e_{\mathrm{lim}}$ = 5 MIP. For the highly granular core of the calorimeter, this corresponds to an energy density of 7.4 GeV/1000 cm$^3$ in Figure \ref{fig:bins_LC} (a). For the energy range studied, the mean hit energy is between 2.7 to 4.7\,MIP. Figure \ref{fig:gl_compfac} shows the distributions of $C_{\mathrm{global}}$ for different energies, demonstrating its energy dependence, originating from the change of the overall hit energy spectrum with changing particle energy. When applying $C_{\mathrm{global}}$ in the energy reconstruction, this dependence has to be corrected for, as discussed below.

The reconstructed energy with global software compensation is obtained in two steps. First, a corrected shower energy is calculated by multiplying the reconstructed energy in the AHCAL and in the TCMT by the factor $C_{\mathrm{global}}$, giving $E_{\mathrm{shower}} = C_{\mathrm{global}} \cdot \left( E_{\mathrm{HCAL}} + E_{\mathrm{TCMT}} \right)$. From this corrected shower energy, the final reconstructed energy with global software compensation for a given event, $E_{\mathrm{GC}}$,  is then obtained from 
\begin{equation}
E_{\mathrm{GC}} = E_{\mathrm{ECAL}}^{\mathrm{track}} + E_{\mathrm{shower}} \cdot P_{\mathrm{global}}(E_{\mathrm{shower}}),
\label{eq:eventEGC}
\end{equation}
where $P_{\mathrm{global}}(E_{\mathrm{shower}})$  is a function which accounts for the energy dependence of the compensation parameters, visible  in Figure \ref{fig:gl_compfac} as the shift of the mean of $C_{\mathrm{global}}$ with energy. This function depends on the corrected shower energy $E_{\mathrm{shower}}$ and is given by a second-order polynomial, $P_{\mathrm{global}}(E_{\mathrm{shower}}) = a_0 + a_1 \cdot E_{\mathrm{shower}} + a_2 \cdot E_{\mathrm{shower}}^2$. The parameters for this function are obtained from a fit of the dependence of the corrected shower energy $E_{\mathrm{shower}}$ on the true deposited energy given by the beam energy corrected for the energy deposited in the ECAL, and are extracted from a training data set extending over the full energy range considered here. They are found to be  $a_0 = 0.982 \pm 0.007$,   $a_1 = 0.0041 \pm 0.0003$~GeV$^{-1}$ and  $a_2 = (-2.2 \pm 0.3) \cdot 10^{-5}$~GeV$^{-2}$. 

The application of the global software compensation technique does not require knowledge of the beam energy, since the energy reconstructed in the HCAL and TCMT is used also in the determination of the correction of the energy dependence of the compensation parameters.

\section{Results}
\label{sc:data}

To evaluate their performance, both software compensation techniques are applied to test beam data and to simulated data. The parameters for the algorithms are determined using test beam data following the training procedures outlined above. 

\subsection{Application of software compensation to test beam data}

When applying the software compensation techniques to test beam data, the energy dependent compensation factors are determined event-by-event using the uncorrected reconstructed energy.  Figure~\ref{fig:enr_dist} shows the distribution of reconstructed energies for the uncorrected reconstruction compared with both software compensation techniques studied. The results are shown for pions with energies of 10 GeV and 80~GeV. In both cases, the software compensation algorithms improve the energy resolution, evidenced by a narrowing of the distributions, while preserving or even improving the Gaussian form of the distributions. The algorithms also bring the mean value of the reconstructed energy closer to the beam energy, resulting in small shifts of the maxima visible in Figure~\ref{fig:enr_dist}.  The mean reconstructed energy with local and global compensation techniques, compared to the uncorrected response without compensation, is shown in Figure \ref{fig:lin_data}  for all energies studied. For both techniques, all points fall within  $\pm$1.5\% of linearity.

The energy resolution before and after compensation is shown in Figure~\ref{fig:res_data}. Good agreement between the  $\pi^{-}$ and  $\pi^{+}$ samples is observed. The energy dependence of the energy resolution is well described by Equation \ref{eq:relres} with a fixed noise term $c$ = 0.18~GeV as discussed in Section \ref{reco:reco}. The fit results are summarised in Table~\ref{tab_fit}. The application of software compensation results in a decrease of the stochastic term while the constant term remains unchanged. Both compensation techniques show very similar performance, with the local software compensation providing a slightly smaller stochastic term, and slightly better performance at intermediate energies. 

\begin{table}[h]
 \caption{Stochastic, constant and noise term contributions to the resolution of the CALICE AHCAL determined with a fit of Equation \protect \ref{eq:relres} to data.}
 \label{tab_fit}
 \begin{center}
  \begin{tabular}{l|c|c|c}
   
& a [\%] & b [\%] & c [GeV]   \\
\hline
\hline
uncorrected & 57.6$\pm$0.4 & 1.6$\pm$0.3 & 0.18  \\
\hline
local compensation &  44.3$\pm$0.3  & 1.8$\pm$0.2 & 0.18 \\
\hline
global compensation &  45.8$\pm$0.3  & 1.6$\pm$0.2 & 0.18  \\
  \end{tabular}
 \end{center}
\end{table}

\begin{figure}
 \begin{center}
  \includegraphics[width=0.7\textwidth]{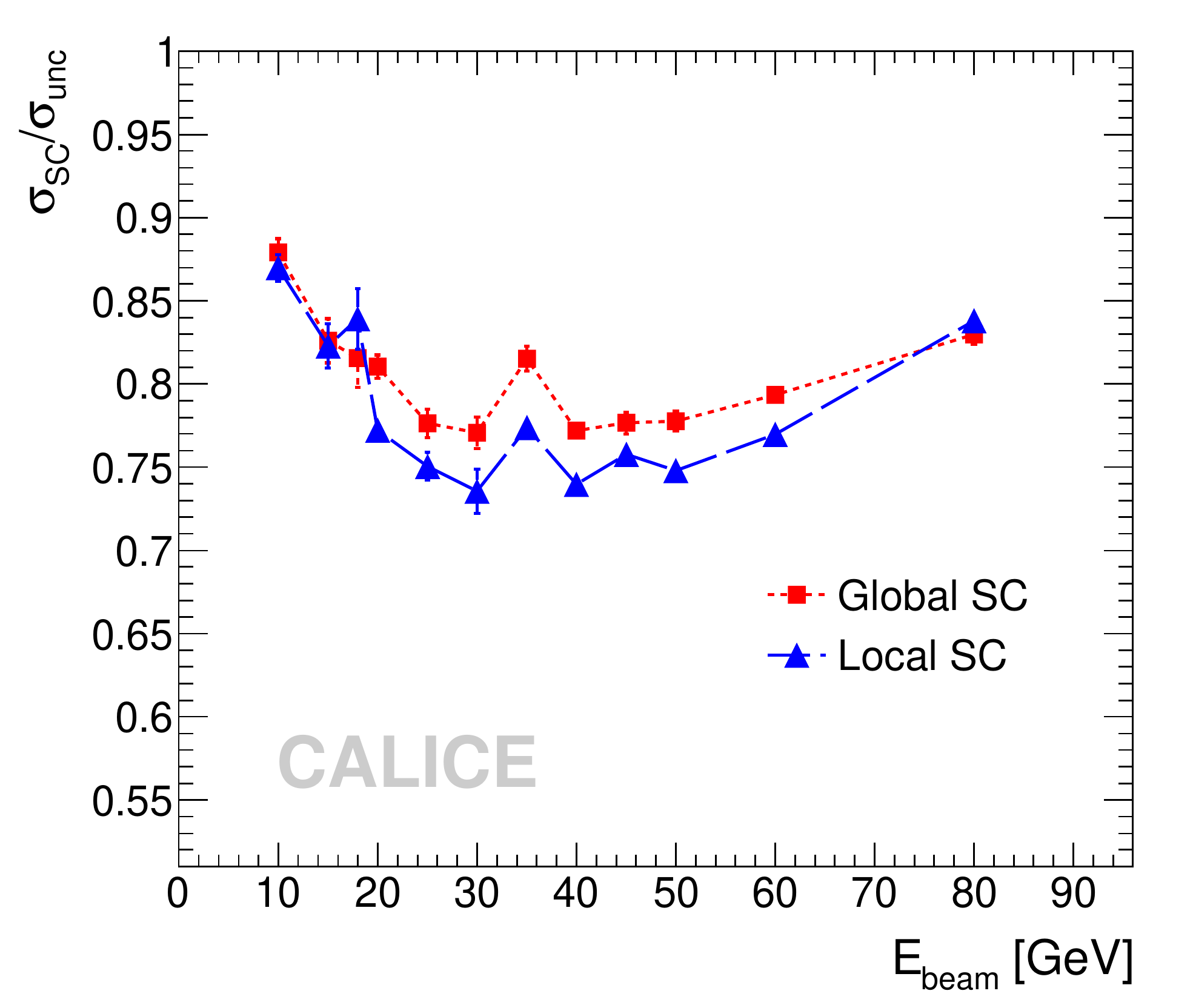}
  \caption{Energy dependence of the relative improvement of the resolution with local and global software compensation observed for data. Where available, results for $\pi^-$ and $\pi^+$ are averaged for clarity.}
  \label{fig:improvementData}
 \end{center}
\end{figure}

Figure \ref{fig:improvementData} shows the relative improvement of the energy resolution achieved with the software compensation techniques, defined as the ratio of the resolution after software compensation $\sigma_{\mathrm{SC}}$ (local or global) to the uncorrected resolution $\sigma_{\mathrm{unc}}$. The improvement ranges from  $\sim$12\% to $\sim$25\% in the energy range studied, for both techniques, with approximately 3\% better relative improvement observed for the local technique in the energy range from 25 GeV to 60 GeV. The reduced performance at high energy is partially due to increased leakage into the TCMT. Energy deposits in the TCMT are not weighted in the local software compensation since their energy density is not well defined. In the global software compensation, the weight is applied also to TCMT energy deposits, but those are not considered in the determination of the weighting factor due to the different readout geometry which leads to increased uncertainties in the weight determination.

\subsection{Comparison to Monte Carlo simulations}
\label{sc:mc}

\begin{figure}
 \begin{center}
  \includegraphics[width=0.7\textwidth]{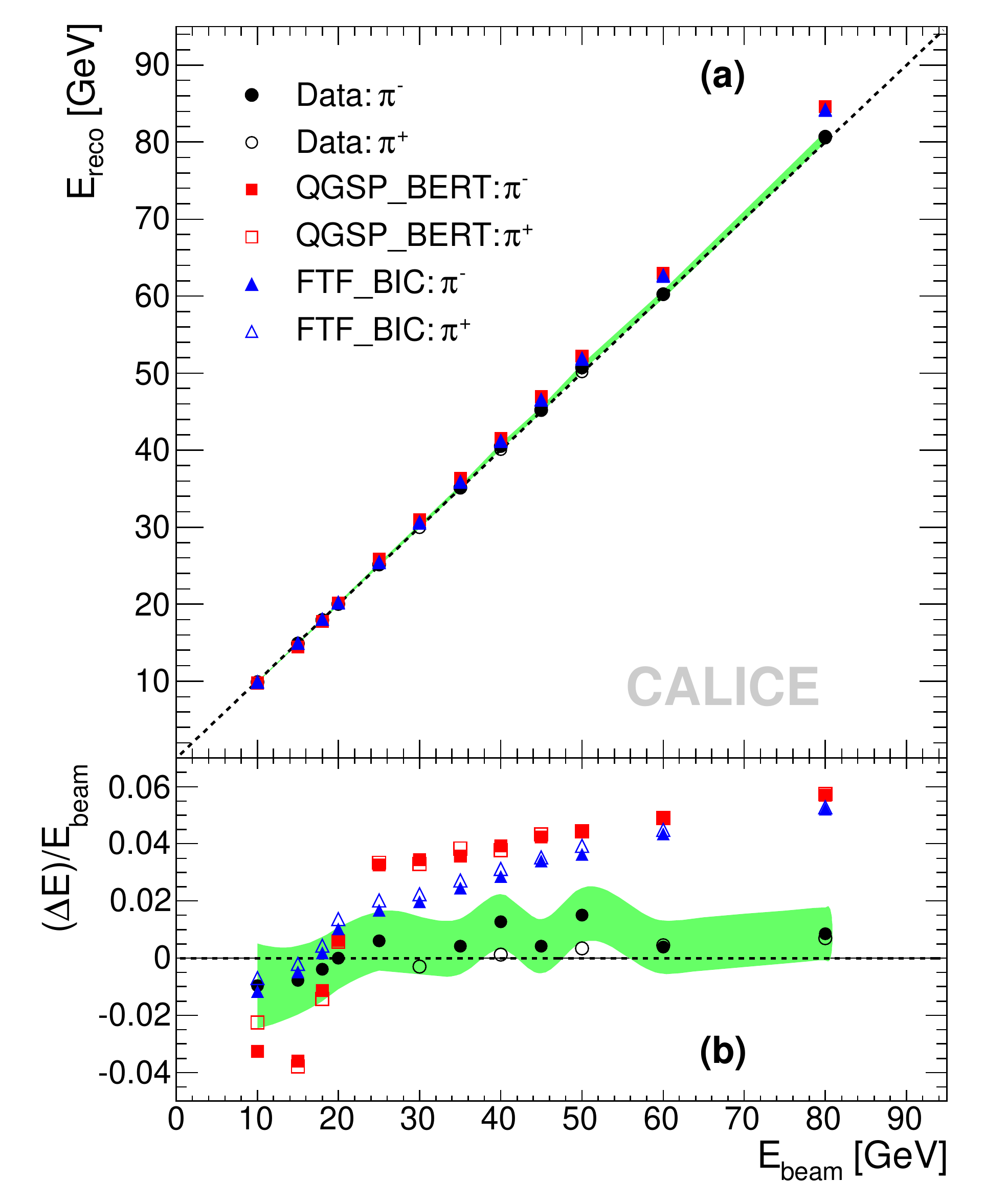}\\
  \caption{(a) Uncorrected response to pions and (b) relative residuals to beam energy versus beam energy for data (black circles), QGSP\_BERT (red squares) and FTF\_BIC (blue triangles). Filled and open markers indicate $\pi^{-}$ and $\pi^{+}$, respectively. Dotted lines correspond to $E_{\mathrm{reco}} = E_{\mathrm{beam}}$, while the green band shows systematic uncertainties for the uncorrected $\pi^{-}$ data sample.}
  \label{fig:linMCini}
 \end{center}
\end{figure}

The stability of both software compensation techniques, as well as the realism of simulation models,  is tested using Monte Carlo simulations. For this purpose, the software compensation algorithms with coefficients derived from data are applied to Monte Carlo samples generated with a detailed detector model in {\sc geant}4.9.4 \cite{Agostinelli:2002hh} using two physics lists: QGSP\_BERT and FTF\_BIC~\cite{Geant4PhysicsLists}. The QGSP\_BERT physics list was chosen because it is the most widely used model in high energy physics experiments at present. The FTF\_BIC physics list, in turn, has provided good results in a previous CALICE analysis \cite{Adloff:2010xj} and is completely independent from QGSP\_BERT.

\begin{figure}
 \begin{center}
  \includegraphics[width=0.7\textwidth]{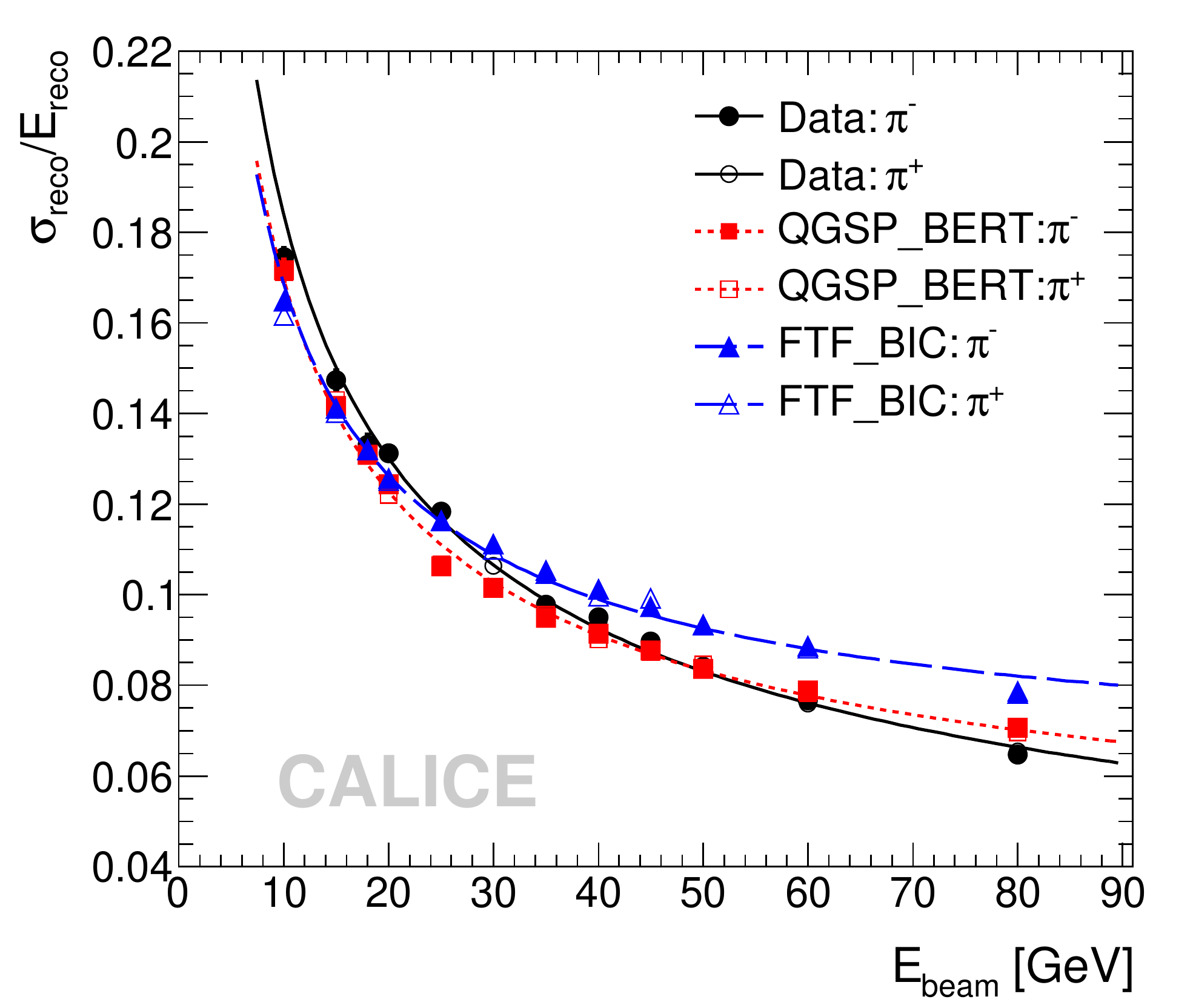}\\
  \caption{Uncorrected energy resolution versus beam energy for data as well as simulations using the physics lists QGSP\_BERT  and FTF\_BIC. The curves show fits using Equation~{\protect \ref{eq:relres}}. The stochastic terms are $(57.6\pm0.4)\%$, $(51.8\pm0.3)\%$ and $(49.4\pm0.3)\%$, with constant terms of $(1.6\pm0.3)\%$, $(4.0\pm0.1)\%$ and $(6.1\pm0.1)\%$ for data, QGSP\_BERT and FTF\_BIC, respectively.}
  \label{fig:resMCini}
 \end{center}
\end{figure}

 Details on the simulation procedure for the AHCAL can be found in \cite{collaboration:2010rq}. For the chosen physics lists, samples of $\pi^{+}$ and $\pi^{-}$ events were simulated at the same energies as the data points using beam profiles, detector temperatures and voltage settings from the data runs. The calibration of the simulation was performed at the MIP level by converting the simulated energy deposits in the scintillator into MIPs using the most probable energy loss of muons determined in simulations. The simulated data sets were passed through the same event selection and reconstruction procedures as real data, using the conversion factors from the MIP scale to reconstructed energy determined for data as discussed in Section \ref{reco:reco}. This also includes the requirement for an identified primary inelastic interaction in the first five layers of the AHCAL.

The uncorrected reconstructed energy as a function of beam energy is shown for data and both physics lists in Figure ~\ref{fig:linMCini} (a). The relative deviation from the beam energy, shown in Figure \ref{fig:linMCini} (b), indicates that simulations with both physics lists behave differently than the data. Both models overestimate the reconstructed energy at high particle energies. In addition, QGSP\_BERT exhibits fluctuations in the transition region between different models in the region between 10~GeV and 20~GeV. In general, the reconstructed energy for simulations is less linear than for data. 

Figure \ref{fig:resMCini} shows the energy resolution without software compensation, comparing data and simulations. Again, the behaviour of the simulations is different from that of the data, with both models underestimating the resolution at low energy, and with FTF\_BIC overestimating the resolution above \mbox{30\,GeV}. This difference leads to a reduced stochastic resolution term with a significantly increased constant term. The results of the fits to the data points using Equation \ref{eq:relres} are summarised in Table \ref{tab:FitMC}.

\begin{figure}
\vspace{1cm}
 \centering
  \includegraphics[width=0.49\textwidth]{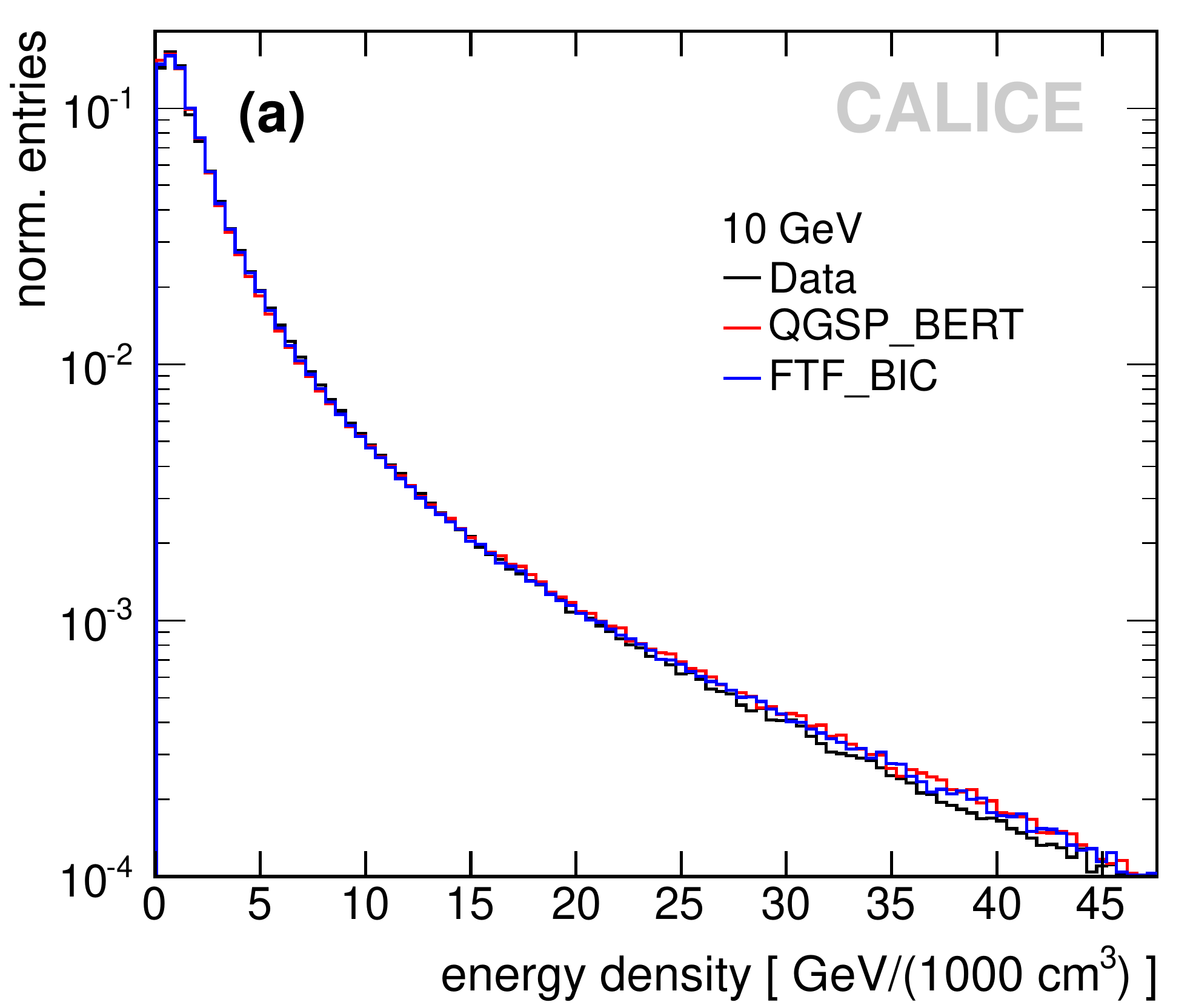}
  \hfill	
  \includegraphics[width=0.49\textwidth]{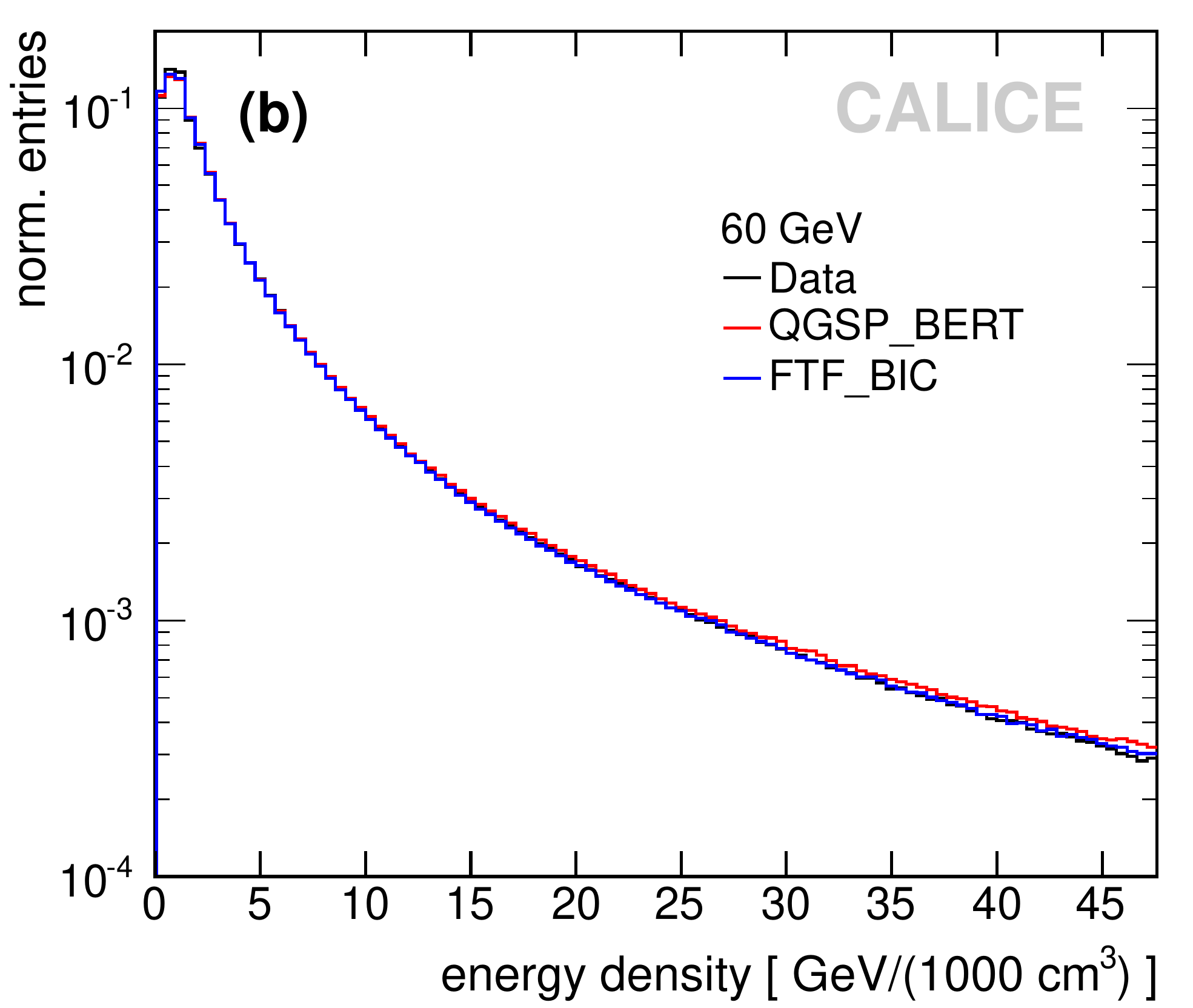}
  \caption{Energy density distribution for data and the two physics lists QGSP\_BERT and FTF\_BIC for (a) 10 GeV and (b) 60 GeV. The energy density is given by the uncorrected reconstructed energy in that cell, divided by the corresponding absorber volume. The distributions are normalized to the number of entries (number of hits).}
  \label{fig:EnergyDensityMC}
\end{figure}

The shape of the distribution of the hit energy density is quite well reproduced by both physics lists, providing the basis for an application of the software compensation algorithms to simulations using the parameters determined from data. Figure \ref{fig:EnergyDensityMC} shows the distributions for two representative energies. The distributions are normalized to the number of entries to show the overall shape while ignoring differences in the normalisation originating from different energy sums. The distributions show a slight overestimation of the fraction of high-density hits by the simulations. In addition, uncertainties in the treatment of saturation effects of the photon sensor lead to an excess of cells with very high energy content well beyond the range shown in the figure. 

\begin{figure}
 \begin{center}
  \includegraphics[width=0.49\textwidth]{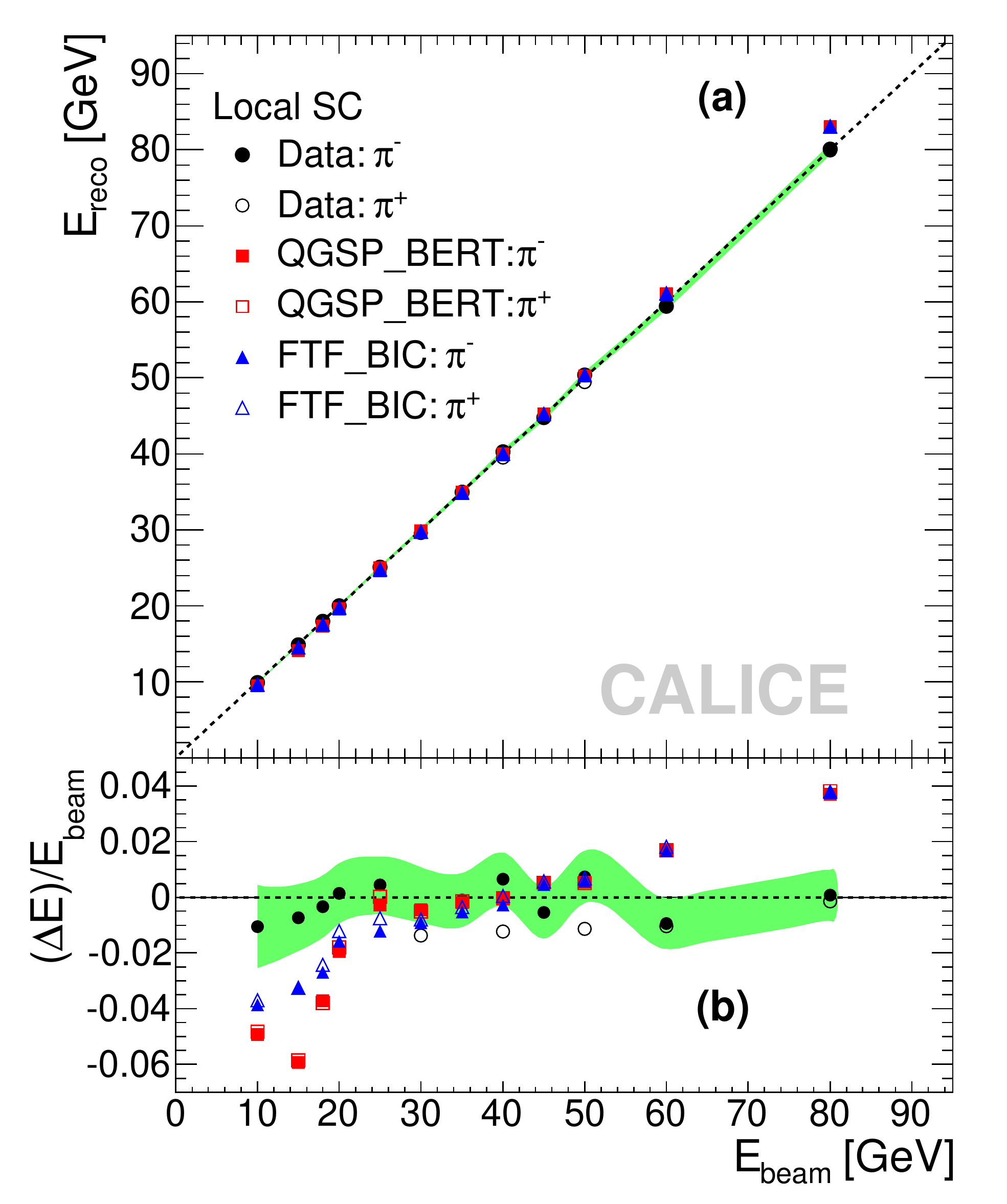}
  \hfill
  \includegraphics[width=0.49\textwidth]{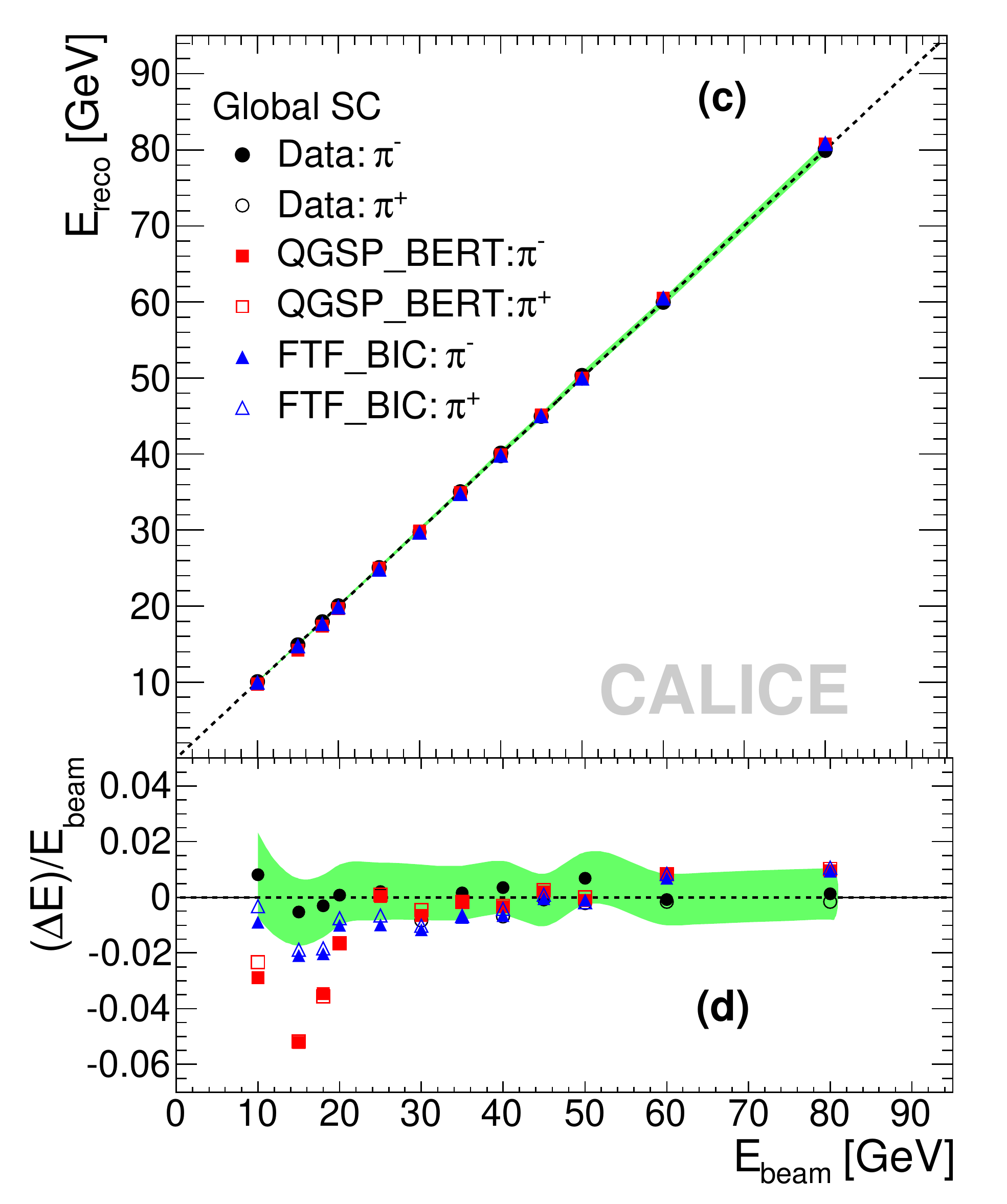}
  \caption{Detector response to pions with software compensation comparing data and simulations. For both data and simulations compensation parameters derived from data are used. (a) Response with local software compensation and (b) corresponding relative residuals to beam energy. (c) Response with global software compensation and (d) corresponding relative residuals to beam energy.}
  \label{fig:linearityMCSC}
 \end{center}
\end{figure}

The effect of the application of the software compensation algorithms, with parameters extracted from data, on the reconstructed energy in  simulations is shown in Figure \ref{fig:linearityMCSC}. For both compensation techniques, the underestimation of the detector response at low energy, in particular by the QGSP\_BERT physics list, remains present. At intermediate energies from 20 GeV up to \mbox{50 GeV}, the application of software compensation results in an improved response linearity and in a better agreement between data and simulations for both physics lists considered. At higher energy, a significant overestimation of the reconstructed energy by simulations is seen with local software compensation, while the global software compensation technique successfully corrects the non-linearity of the simulations in that energy regime. This difference in behaviour is partially due to uncertainties in the treatment of saturation effects in simulations, and potentially also receives a contribution from imperfect descriptions of the shower structure by the shower models themselves. In the simulations, the number of cells with very high energy content is overestimated and exhibits a longer tail than in data, as discussed above. This affects the correction factor of the global software compensation by construction, resulting in a lower shower weight for simulations compared to data at the same energy on average, bringing data and simulations into better agreement. The local software compensation technique applies constant weights for very high-energy hits, as can be seen in Figure \ref{fig:bins_LC} (b). It is thus less sensitive to these differences between data and simulations and preserves the discrepancy in visible energy for high beam energies.

\begin{figure}
 \begin{center}
  \includegraphics[width=0.49\textwidth]{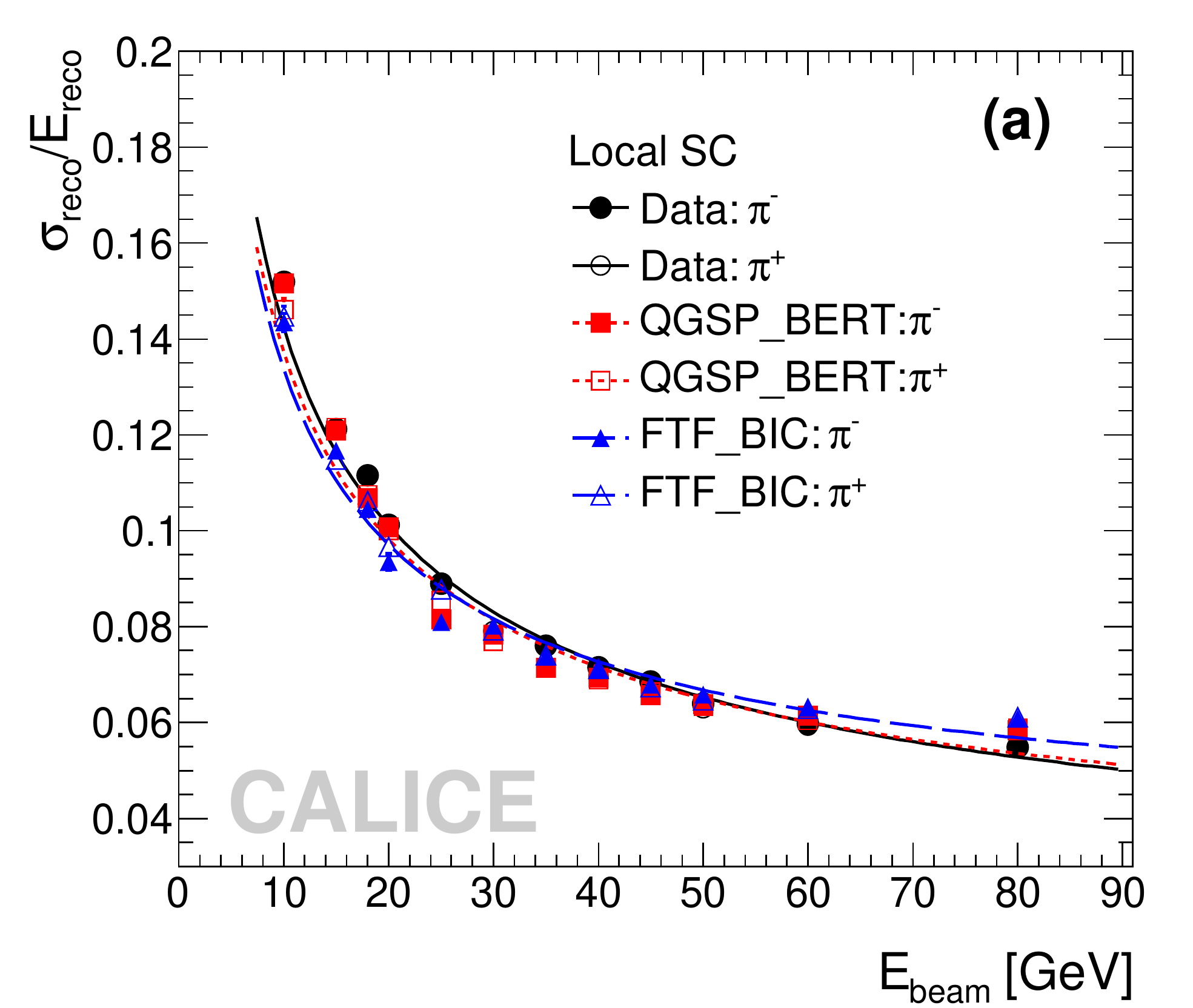}
  \hfill
  \includegraphics[width=0.49\textwidth]{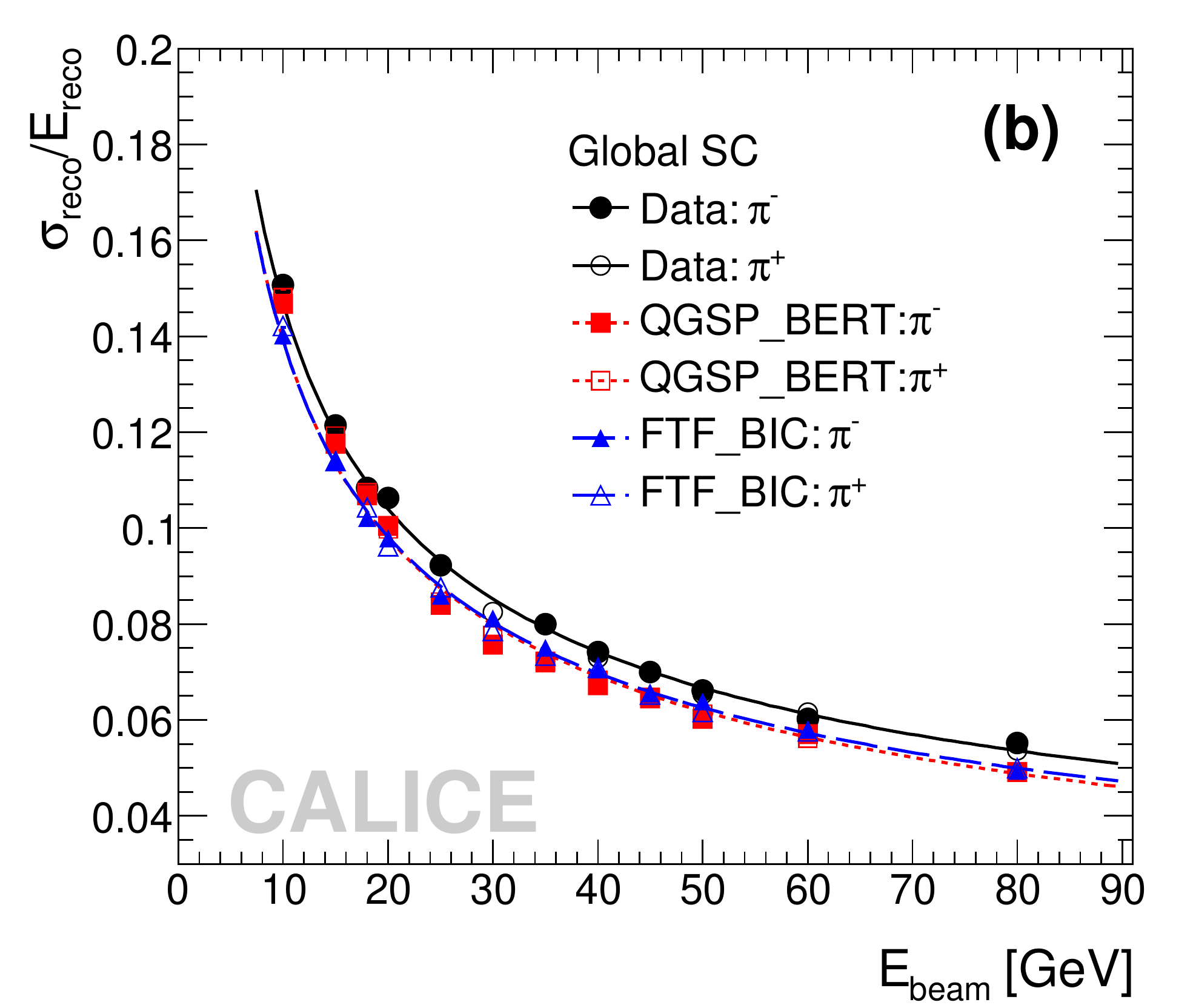}
  \caption{Energy resolution for pions with local (a) and global (b) software compensation comparing data and simulations. For both data and simulations compensation parameters derived from data are used. The curves show fits using Equation~{\protect \ref{eq:relres}}. The fit results for the local software compensation are  $(44.3\pm0.3)\%$, $(42.3\pm0.2)\%$ and $(40.4\pm0.3)\%$ for the stochastic term, with constant terms of $(1.8\pm0.2)\%$, $(2.5\pm0.1)\%$ and $(3.4\pm0.1)\%$ for data, QGSP\_BERT and FTF\_BIC, respectively. For the global software compensation, the results are  $(45.8\pm0.3)\%$, $(43.6\pm0.2)\%$ and $(43.4\pm0.3)\%$ for the stochastic term, with constant terms of $(1.6\pm0.2)\%$, $(0.0\pm0.2)\%$ and $(1.1\pm0.2)\%$ for data, QGSP\_BERT and FTF\_BIC, respectively. }
  \label{fig:resolutionMCSC}
 \end{center}
\end{figure}

\begin{table}
 \caption{Fit results using the function given in Equation (\protect \ref{eq:relres}) for simulations with and without software compensation, compared to the corresponding values for data.}
 \label{tab:FitMC}
 \begin{center}
  \begin{tabular}{l|c|c|c}

 & a [\%] & b [\%] & c [GeV] \\
\hline
\hline
uncorrected data & 57.6$\pm$0.4 & 1.6$\pm$0.3 & 0.18   \\
uncorrected QGSP\_BERT & 51.8$\pm$0.3 & 4.0$\pm$0.1 & 0.18  \\
uncorrected FTF\_BIC & 49.4$\pm$0.3 & 6.1$\pm$0.1 & 0.18 \\
\hline
local compensation data &  44.3$\pm$0.3  & 1.8$\pm$0.2 & 0.18   \\
local compensation QGSP\_BERT &  42.3$\pm$0.2  & 2.5$\pm$0.1 & 0.18  \\
local compensation FTF\_BIC &  40.4$\pm$0.3  & 3.4$\pm$0.1 & 0.18  \\
\hline
global compensation data &  45.8$\pm$0.3  & 1.6$\pm$0.2 & 0.18   \\
global compensation QGSP\_BERT &  43.6$\pm$0.2  & 0.0$\pm$0.2 & 0.18  \\
global compensation FTF\_BIC &  43.4$\pm$0.3  & 1.1$\pm$0.2 & 0.18   \\
  \end{tabular}
 \end{center}
\end{table}

Figure \ref{fig:resolutionMCSC} shows the energy resolution for simulations compared to that for data for both software compensation techniques. The local software compensation largely preserves the differences between data and simulations for the physics list QGSP\_BERT, but results in a better agreement of FTF\_BIC with data, in agreement with the behaviour observed for the reconstructed energy. The global software compensation brings the overall trend of the resolution with energy for data and simulations into good agreement, with better resolution seen for simulations with both physics lists than for data. The results of the fits to the data points using Equation \ref{eq:relres} are summarised in Table~\ref{tab:FitMC}, together with the results obtained without software compensation.

\begin{figure}
 \begin{center}
  \includegraphics[width=0.99\textwidth]{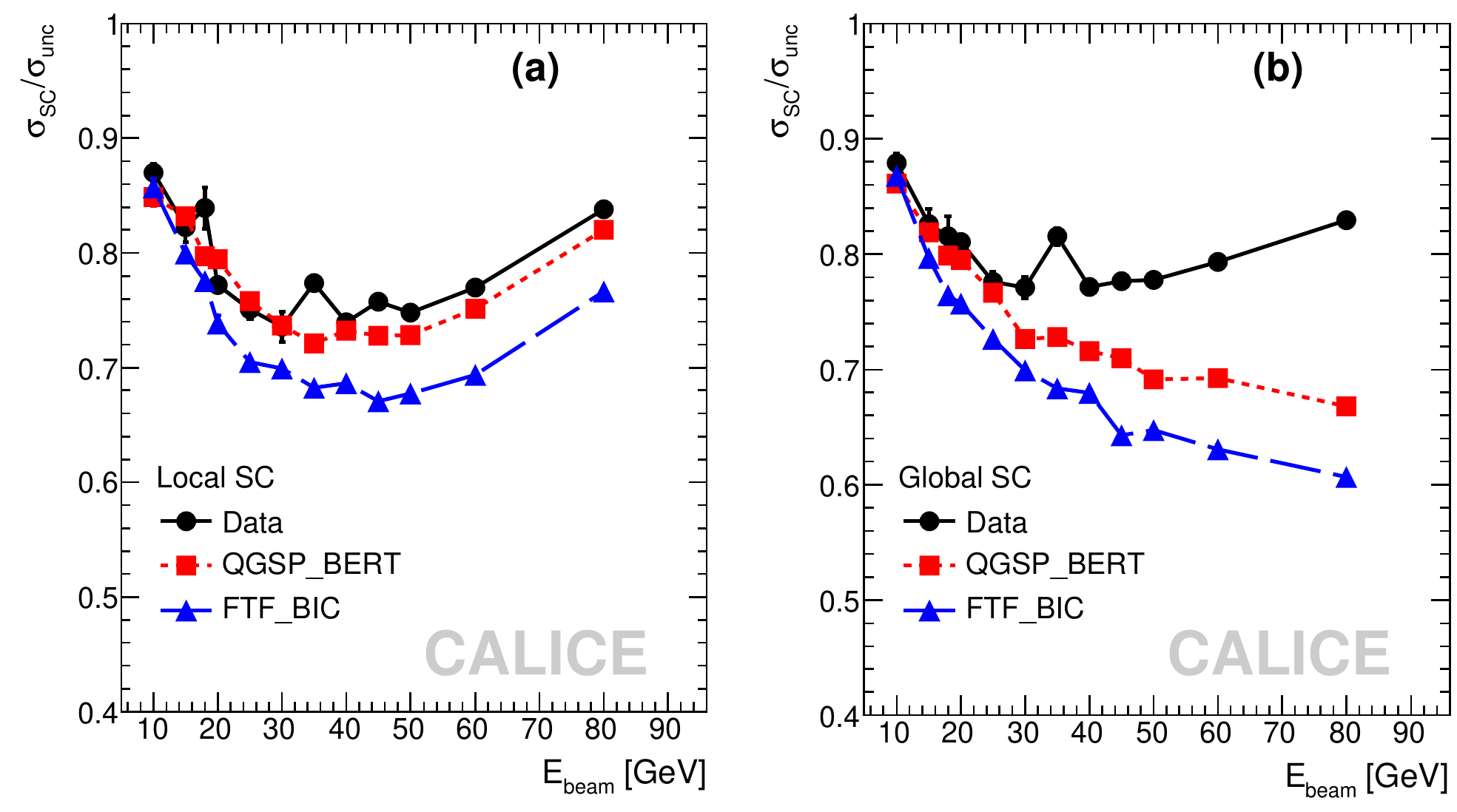}\\
  \caption{Energy dependence of the relative improvement of the resolution for data and simulations using the physics lists QGSP\_BERT  and FTF\_BIC, (a) with local software compensation and (b) with global software compensation. Where available, results for $\pi^-$ and $\pi^+$ are averaged for clarity.}
  \label{fig:improvementMC}
 \end{center}
\end{figure}

The relative improvement in resolution compared to the uncorrected energy resolution is shown in Figure \ref{fig:improvementMC} for data and simulations. For the local software compensation, the improvement with respect to energy observed in data is well reproduced by the QGSP\_BERT physics list. For FTF\_BIC, a considerably bigger improvement is seen for the simulations at high energy than is seen in data. This higher improvement at high energies results in the better agreement of the energy resolution in data and in simulations discussed above. 
For the global compensation approach, the behaviour up to 30 GeV is well modelled by QGSP\_BERT, while an up to 20\% higher improvement, compared to that for data,  is seen in simulations at the highest energies considered. The reason for this different behaviour of local and global software compensation is the same as for the different high-energy behaviour for the reconstructed energy as discussed above.

\section{Conclusion}

The hadronic energy resolution of the  CALICE analogue hadron calorimeter is studied using test beam data collected in 2007 at the CERN SPS. The calorimeter, with an instrumented volume of approximately 1~m$^3$ and a depth of 5.3~$\lambda_I$, is highly segmented in both longitudinal and lateral direction, with a total of 7608 electronic channels. The intrinsic energy resolution of the CALICE AHCAL for hadrons is measured to be $ 58\%/\sqrt{E/\mathrm{GeV}}$, with a constant term of 1.6\%. 

The unprecedented granularity of the CALICE AHCAL provides excellent possibilities for the application of software compensation algorithms to improve the energy resolution of the calorimeter based on event-by-event information on the energy density structure of the showers. Two techniques have been presented here, together with results from test beam and from simulated data samples. The local software compensation technique uses local energy density information for a cell-by-cell re-weighting of energy deposits, while the global software compensation technique uses the distribution of cell energies to derive one overall weighting factor for each shower. Both techniques show similar performance, with a relative improvement of the energy resolution ranging from 12\% to 25\% over the studied energy range from 10 GeV to 80 GeV, resulting in a reduction of the stochastic term to $45\%/{\sqrt{E/\mathrm{GeV}}}$. In {\sc geant}4 simulations with the QGSP\_BERT and the FTF\_BIC physics lists, the detector response is considerably more non-linear than in data. The physics list QGSP\_BERT provides a satisfactory description of the energy resolution. The application of software compensation using parameters determined from data brings the resolution into better agreement with data. Here, the improvement of the energy resolution using the local software compensation technique observed for the QGSP\_BERT physics lists is comparable to that observed for data,  while larger differences are observed for FTF\_BIC and for the global software compensation technique. 

Neither of the described techniques requires an {\it a priori} knowledge of the particle energy. The energy dependent compensation factors are selected based on the uncorrected reconstructed energy. Although this energy dependence places some restrictions on the implementation of both techniques in a collider environment with a high particle density in hadronic jets, their application in the context of particle flow algorithms should be possible based on identified calorimeter clusters. The jet energy resolution can profit from the improved hadronic energy resolution directly through a better measurement of the neutral hadronic component, but also indirectly from an improved matching of reconstructed tracks and calorimeter energy during the clustering phase. 

\acknowledgments

We gratefully acknowledge the DESY, CERN and Fermilab managements for their support and
hospitality, and their accelerator staff for the reliable and efficient
beam operation. 
We would like to thank the HEP group of the University of
Tsukuba for the loan of drift chambers for the DESY test beam.
The authors would like to thank the RIMST (Zelenograd) group for their
help and sensors manufacturing.
This work was supported by the 
Bundesministerium f\"{u}r Bildung und Forschung, Germany;
by the  the DFG cluster of excellence `Origin and Structure of the Universe' of Germany ; 
by the Helmholtz-Nachwuchsgruppen grant VH-NG-206;
by the BMBF, grant no. 05HS6VH1;
by the Alexander von Humboldt Foundation (Research Award IV, RUS1066839 GSA);
by joint Helmholtz Foundation and RFBR grant HRJRG-002, SC Rosatom;
by Russian Grants  SS-1329.2008.2 and RFBR08-02-121000-0FI
and by the Russian Ministry of Education and Science contract 02.740.11.0239;
by MICINN and CPAN, Spain;
by CRI(MST) of MOST/KOSEF in Korea;
by the US Department of Energy and the US National Science
Foundation;
by the Ministry of Education, Youth and Sports of the Czech Republic
under the projects AV0 Z3407391, AV0 Z10100502, LC527  and LA09042  and by the
Grant Agency of the Czech Republic under the project 202/05/0653;  
and by the Science and Technology Facilities Council, UK.

\end{document}